\documentstyle[preprint,eqsecnum,aps]{revtex}
\font\eu=eufm10
\begin{document}
\draft
\tightenlines

\title{LENZ-ISING-ONSAGER PROBLEM IN AN EXTERNAL FIELD AS A 
SOLUBLE PROBLEM OF MANY FERMIONS}

\author{Martin S. Kochma\'nski}
\address{Institute of Physics, 
Pedagogical University\\
T.Rejtana 16 A , 35--310 Rzesz\'ow, Poland\\
e-mail: mkochma@atena.univ.rzeszow.pl}
\maketitle

\begin{abstract}
In this paper a new approach to solving the $2D$ and $3D$ Ising 
models in external magnetic field $H\neq0$ is developed. The 
general formalism for the approach to the problem is presented on the example 
of the $2D$ Ising model in the external magnetic field. The paper presents a 
new method obtaining the Onsager solution and computations of asymptotic 
forms of low-temperature free energy for the $2D$ Ising model in the external 
magnetic field $(H)$. The free energy in the limiting case of the magnetic
field  tending to zero $(H\rightarrow 0, \;\;\; N,M\rightarrow\infty)$ 
at arbitrary temperature is also considered $(T\neq 0)$.
\end{abstract}

\widetext
\section{Introduction}
\label{sec: level1}

We will briefly describe the well known model of a magnetic containing(?) 
variety of spins situated on the vertices of a crystalline lattice. The spin 
at $k$ can be "up" $(\sigma_k=1)$ or "down" $(\sigma_k=-1)$. 
A microscopic state of the system is characterized by orientations of all the 
spins. Energy $E\{\sigma\}$ of the microscopic 
state $\{\sigma\}$ is composed of two contributions, one from the exchange 
interactions of the spins and described by the interaction constant $J_{kl}$, 
and the second from the interaction of the spins with the external magnetic 
field $(H)$:

\begin{equation}
E\{\sigma\}=-\sum_{kl}J_{kl}\sigma_{k}\sigma_{l}-H\sum_{k}\sigma_{k},
\end{equation}
where summation is taken over all sites of the lattice. The key problem is 
calculation of the statistical sum:
\begin{equation}
Z=\sum_{\{\sigma_{k}\}}\exp(-\beta E\{\sigma\}), \;\;\;\;\;\;\;\;\;
\beta =\frac{1}{k_{B}T},
\end{equation}
where $T$ denotes temperature and $k_{B}$ - the Boltzman constant.

The model described above was introduced by W.Lenz in $1920$ $\cite{lenz20}$, 
and for the one-dimensional case was investigated by E.Ising in $1925$ 
$\cite{ising25}$. The first exact solution of the statistical mechanical 
problem for the $2D$, $(H=0)$ case was found by L.Onsager in $1944$ 
$\cite{onsager44}$. We use the standard name, the Ising model.

The solution given by Onsager strongly influenced the development of all of 
statistical physics, and in particular of the theory of phase transitions. 
It was shown for the first time that exact calculation of the free energy 
leads to an evidence that thermodynamic quantities behave in the vicinity of 
the phase transition in a way which is essentially different from that in the 
approximate models, like e.g. the mean-field theory. The result for 
spontaneous magnetization {\eu M$_0$} in the model was presented by 
Onsager at the conference in Florence in $1949$ $\cite{onsager49}$, i.e. $5$ 
years after the successful derivation of the expression for free energy. The 
first published derivation for {\eu M$_0$} was given by Yang 
$\cite{yang52}$, are recently, alternative derivations have been published 
both for the free energy and {\eu M$_0$} $\cite{sml64,baxter82,izyum87,mccoy-wu73}$.

In spite of its simplicity, the Ising model is not only non-trivial in higher 
dimensions $(d\geq 2)$, but also it has rich structure. By this we mean not 
only its connection with other models (for example with the lattice gas 
models, binary alloys, some models in quantum field theory $\cite{baxter82,gaudin83}$ 
etc.), and wide application in numerous domains of statistical physics, but 
also its role as a generator of new ideas and tools, which find its 
use in various areas of physics and mathematics. There are sufficiently many 
examples of such applications and we will not discuss them here (some examples 
can be found in the monograph $\cite{liggett85}$, where stochastic Ising 
models are considered, and also their connection with Markov processes with 
local interactions). We would like to stress that this  
rich structure of the Ising model has maintained a high level of  
interest in this problem among physicists and mathematicians.

In this paper we present a new approach to the Ising problem in external 
magnetic field $(H)$, with the nearest-neighbour interaction on the square 
lattice. In connection with that we would like to mention the paper by Schultz, 
Mattis and Lieb $\cite{sml64}$, who applied it to solve the $2D$ Ising model 
without an external magnetic field. To calculate {\eu M$_0$} they 
used a method based on a transfer-matrix using a transformation 
to a fermionic representation. This deep, clear and logically 
closed paper influenced strongly the author and moved him to search for the 
solution of the problem in external magnetic field. The fundamental idea of 
the approach of the authors of the paper $\cite{sml64}$ is transition to a 
fermionic representation (the transfer-matrix method was essentially used 
already in the paper by Onsager $\cite{onsager44}$), and this can be treated 
in a sense as a problem of interacting fermions on the 
one-dimensional lattice. In this paper we use essentially the same idea. The 
difference is the fermionic representation is introduced not on a $1D$ 
lattice (where the $T$-matrix is expressed in terms of the Fermi creation and 
annihilation operators $(c^{\dag}_{n}, c_{n})$, $\cite{sml64}$) but on a 
two-dimensional lattice with the doubly indexed Fermi creation - annihilation 
operators $(c^{\dag}_{nm},c_{nm})$, $\cite{mkoch95}$.

\section{Formulation of the problem}

Let us consider the square lattice composed of $M$ columns and $N$ rows, on 
the vertices of which the quantities $\sigma_{nm}$ taking one of the two values 
$\pm 1$ are defined. We will call the quantities the Ising "spins". The 
multiindex $nm$ numbers the sites of the lattice, where $n$ numbers a row, and 
$m$ numbers a column. The Ising model with the nearest-neighbour interaction 
in external magnetic field is given by the Hamiltonian of the form:
\begin{equation}
{\cal H}=-J_{2}\sum_{nm}\sigma_{nm}\sigma_{n+1,m}-J_{1}\sum_{nm}
\sigma_{nm}\sigma_{n,m+1}-H\sum_{nm}\sigma_{nm} , 
\end{equation}
which takes into account anisotropy in the interaction $(J_{1,2}>0)$ between 
nearest neighbours, and also the interaction of the spins $\sigma_{nm}$ with 
external magnetic field $H$, directed "up" $(\sigma_{nm}=+1)$. The essential 
problem is calculation of the statistical sum for the system:
\begin{eqnarray*}
Z(h)=\sum_{\sigma_{11}=\pm 1}...\sum_{\sigma_{NM}=\pm 1}\exp\left(-
\beta{\cal H}\right)
\end{eqnarray*}
\begin{equation}
=\sum_{(\sigma_{nm}=\pm 1)}\exp\left[\sum^{NM}_{n,m=1}(K_{2}\sigma_{nm}
\sigma_{n+1,m}+K_{1}\sigma_{nm}\sigma_{n,m+1}+h\sigma_{nm})\right] ,
\end{equation}
where 
\begin{equation}
K_{1,2}={\beta}J_{1,2}, \;\;\;\;\; h={\beta}H, \;\;\;\; \beta=1/k_{B}T .
\end{equation}
Periodic boundary conditions are introduced for the variables 
$\sigma_{nm}$. Let us mention here that the statistical sum $(2.2)$ is 
symmetric with respect to the change $h\rightarrow -h$ where $h$ is defined 
above $(2.3)$.

As is known $\cite{sml64}$, the statistical sum for the $2D$ Ising model in 
external field $(H)$ in the representation of second quantization can be 
written in the form:
\begin{equation}
Z=Tr(V)^{N}=Tr(V_{1}V_{2}V_{h})^{N},
\end{equation}
where the operators $\it V_{i}$, expressed in terms of the Fermi creation and 
annihilation operators ($c^{\dag}_{m}, c_{m}$), are of the form: 
\begin{eqnarray}
V_{1}=(2{\sinh}2K_{1})^{M/2}\exp{\left[-2K_{1}^{*}\sum_{m=1}^{M}(c_{m}^{\dag}
c_{m}-1/2)\right]},
\end{eqnarray}
\begin{eqnarray}
V_{2}=\exp\left\{K_{2}\left[\sum_{m=1}^{M-1}(c_{m}^{\dag}-c_{m})(c_{m+1}^{\dag}+
c_{m+1}) -(-1)^{\hat{M}}(c_{M}^{\dag}-c_{M})(c_{1}^{\dag}+c_{1})\right]\right\},
\end{eqnarray}
\begin{eqnarray}
V_{h}=\exp\left\{h\sum_{m=1}^{M}\exp\left[i\pi\sum_{p=1}^{m-1}
c_{p}^{\dag}c_{p}\right]
(c_{m}^{\dag}+c_{m})\right\},
\end{eqnarray}
where $K_{j} , (j=1,2,)$ and $\it h$ are defined above  $(2.3)$ and 
$\hat{M}=\sum_{1}^{M}c_{m}^{\dag}c_{m}$ is the operator of the total number of 
particles and $K_{1}^{*}$ and  $K_{1}$ are connected by the following formulas:
\begin{equation}
\tanh(K_{1})=\exp(-2K_{1}^{*}), \;\;\; or \;\;\; \sinh2K_{1}\sinh2K_{1}^{*}=1.
\end{equation}
One can see that the operator $V_h$ in the second quantization representation, 
that describes interaction of the spins with external magnetic field, has 
rather complicated structure. It is easy to see that this operator does not 
commute with the operator $\hat{P}\equiv(-1)^{\hat{M}}$. As a result the 
operator $V_2$ has also not a very tractable form, i.e. it has not the needed 
translational symmetry $(2.6)$. More exactly, although  the operators $V_1$ 
and $V_2$ commute with the operator $\hat{P}$, the operator $V$ $(2.4)$ does 
not commute with the operator $\hat{P}$, i.e. $[\hat{P},V]_{-}\neq{0}$, 
because $[\hat{P},V_h]_{-}\neq{0}$. Therefore, we can not divide all states 
of the operator $V=V_{1}V_{2}V_h$ into eigenstates of the operator $\hat{P}$ 
with eigenvalues $\lambda=\pm 1$, and this leads to nonconservation of the 
states with even and odd numbers of fermions (for details see $\cite{sml64}$). 
Namely this is the fundamental reason which stops solving the problem under 
consideration within this formalism. Nevertheless, the author does not share 
Ziman's pessimism $\cite{ziman79}$ which is based on some misunderstanding, 
because he considers actually the approach of the authors of the paper 
$\cite{sml64}$, but in the end he writes about limitations of the method of 
Onsager $\cite{onsager44}$. In fact Onsager in his approach does not apply 
the field theoretic language of the creation and annihilation operators as it 
is in the approach of the authors of the paper $\cite{sml64}$. the method of 
Onsager $\cite{onsager44,onkauf49}$ really shows some limitations when one 
tries to apply it to solving the $2D$ Ising model in external magnetic field, 
or for solving the $3D$ Ising model. On the other hand, completely different 
state of affairs we have for the approach of the authors of the paper 
$\cite{sml64}$, where in all its beauty the field theoretic language of the 
method of second quantization is used. The approach of the authors of the 
paper $\cite{sml64}$ allows for generalizations. We intend to present one of 
such generalizations in this and the following papers devoted to the Ising 
problem. 

Coming back to the difficulties mentioned above which are connected with the 
operator $V_h$, $(2.7)$, it is now clear that to overcome the troubles within 
the approach $\cite{sml64}$, one should find an appropriate method of 
substituting the operator $V_h$ $(2.7)$ with another one which would be 
equivalent to the former in the sense of correct counting of the interaction 
of external magnetic field with the spins of the system. Namely, as it could 
be easily seen, the only contribution to $Z$ $(2.4)$ from the operator $V_h$ 
comes, in the representation of second quantization, from the "even" part 
with respect to operators $c^{\dag}_{m},c_{m}$ of the operator $V_h$. In principle 
such transformation could be always done. However, in practice this task 
seems to be hopeless, and the direct method of calculation of the commutators 
used by Onsager for solution of the problem without external field here is 
simply inapplicable. We believe there is not an effective method to do that 
at least if one stays in the space of given dimension ($d=2$ for the initial 
variables $\sigma_{nm}$, and $d=1$ for the variables in the representation of 
second quantization $c^{\dag}_m$ and $c_m$). Nevertheless, as we will show below, 
there is an effective method of transforming the magnetic operator $V_h$, 
$(2.7)$, after which the transformed operator allows for the Fourier transform 
of the operator $V$ $(2.4)$. The idea consists of formulating the problem in 
the space of higher dimension than the former one, then to pass to the 
representation of second quantization with the operator $V$, and afterwards 
to perform a limit transition with respect to one of the interaction constants, 
by going with it to zero. Having this done a possibility appears for effective 
rebuilding of the operator which is responsible for interaction of the spins 
of the system with external magnetic field. Below we will shortly present this 
approach on an example of the one-dimensional Ising model which is then 
applied to solving our basic problem.

\section{One-dimensional Ising model}

In the beginning of the consideration of the $1D$ Ising model we have already 
the complete set of formulas $(2.4-8)$. To apply them to the $1D$ Ising model 
one should take simply $K_1=0$ and $N=1$. Then, after not complicated 
transformations, taking into account the expressions $(2.8)$, one can write 
the following formula for the statistical sum $(2.4)$ [$Z(K_{1}=0)=Z^{*}$]:
\begin{equation}
Z^{*}=Tr(V_{1}^{*}V_{2}V_{h}),
\end{equation}
where the operators $V_{2}$ and $V_{h}$ are defined above $(2.6-7)$, and the 
operator $V_{1}^{*}$ is of the form:
\begin{equation}
V_{1}^{*}=\prod_{m=1}^{M}\left[1+(-1)^{c_{m}^{\dag}c_{m}}\right].
\end{equation}

Introducing in an appropriate manner the basis in the representation of 
occupation numbers [(finite dimensional Fock space): $|0>$ is the vacuum 
state, $c_{m}|0>=0$; $c_{m}^{\dag}|0>$ is a one-particle state $(m=1,2,3,...)$ 
etc.], and calculating the trace in $(3.1)$ we get
\begin{equation}
Z^{*}=\sum_{all(\it l)}<l|(V_{1}^{*}V_{2}V_{h})|l>=2^{M}<0|(V_{2}V_{h})|0>,
\end{equation}
where on the left hand side of $(3.3)$ the summation is over all states $|l>$. 
It is easy to see that all the matrix elements $<l|(...)|l>$ in $(3.3)$ are 
equal to zero thanks to the phase factors $(-1)^{c^{\dag}_{m}c_{m}}$ entering 
the operator $V_1^*$, with the exception of the vacuum matrix element 
$<0|(...)|0>$. For this matrix element contribution from the operator $(3.2)$ 
is equal simply $2^M$. From this we obtain the right hand side of the equality 
$(3.3)$.

Let us mention now that the operators $V_h$, $(2.7)$, can be represented in 
the form: 
\begin{equation}
V_h=\cosh^{M}(h)\prod^{M}_{m=1}\left[1+\psi_{m}(c^{\dag}_{m}+c_{m})\tanh h\right],
\end{equation}
where the phase factor $\psi_{m}$ is defined in an obvious way $(2.7)$ and we 
applied the following identity
\begin{eqnarray*}
\exp(\rho t)=\cosh t+\rho\sinh t, \;\;\;\;\; \rho^{2}=1 .
\end{eqnarray*}
Now, "dragging" the operator $V_{h}$ , $(3.4)$ through the ket-vector $|0>$, 
after a number of transformations, we obtain the following representation 
for $V_h\mid 0>$:
\begin{equation}
V_{h}\mid0>=\cosh^{M}(h)\prod_{m=1}^{M}e^{{\alpha}c_{m}^{\dag}}\mid0>,\;\;\;
\alpha\equiv\tanh(h).
\end{equation}
Deriving the formula $(3.5)$ we dragged all phase factors $\psi_{m}$ entering 
the operator $V_h$ through the vacuum state $|0>$ and omitted the annihilation 
operators $c_m$, because $c_{m}\mid0>=0$. We will omit below, for brevity,  
the ket-vector $|0>$. This should not lead to misunderstandings. Further we 
not that the operators $c_m^+$ and $c_k^+$ commute with the commutator 
$[c_{m}^{\dag},c_{k}^{\dag}]=2c_{m}^{\dag}c_{k}^{\dag}$. As a result, using the Hausdorff-
Baker formula ($\alpha,\beta=const$):
\begin{equation}
\exp({\alpha}x)\exp({\beta}y)=\exp({\alpha}x+{\beta}y+({\alpha}{\beta}/2)
[x,y]),
\end{equation}
after $(M-1)$ times application of this formula $(3.6)$ to the operator $(3.5)$, 
this operator can be represented in the form:
\begin{equation}
V_{h}=\cosh^{M}(h)\exp\left[{\alpha}\sum_{m=1}^{M}c_{m}^{\dag}\right]
\exp\left[{\alpha}^{2}\sum_{m=1}^{M}\sum_{p=1}^{M-m}c_{m}^{\dag}c_{m+p}^{\dag}\right],
\end{equation}
where ${\alpha}$ is defined above $(3.5)$. Since all terms in the operator 
$V_2$, $(2.6)$ contain bilinear products of the Fermi operators, and the 
following equality is satisfied 
\begin{eqnarray*}
\exp\left({\alpha}\sum_{m=1}^{M}c_{m}^{\dag}\right)=1+{\alpha}\sum_{m=1}^{M}c_{m}^{\dag} ,
\end{eqnarray*}
it is easy to see that in the pairings $<0\mid(...)\mid0>$ the components 
linear in $c_m^{\dag}$ give null contribution. As a result, we can write the 
following expression ($V_{h}{\rightarrow}V_{h}^{*}$) for the operator $V_h$, 
$(3.7)$:
\begin{equation}
V_{h}^{*}=\cosh^{M}(h)\exp\left[{\alpha}^{2}\sum_{m=1}^{M}\sum_{p=1}^{M-m}
c_{m}^{\dag}c_{m+p}^{\dag}\right].
\end{equation}
Now one can easily see tat the operator ${\hat {P}}=(-1)^{\hat {M}}$, 
(${\hat {M}}=\sum_{1}^{M}c_{m}^{\dag}c_{m}$) commutes with the operators $V_2$ 
and $V^*_h$, and, as a consequence, the states with even or odd numbers of 
fermions are conserved. Hence, the statistical sum $Z^*$, $(.3)$, can be 
represented in the form:
\begin{equation}
Z^{*}=2^{M}<0\mid(V_{2}^{\pm}V_{h}^{*})\mid0>,
\end{equation}
where
\begin{equation}
V_{2}^{\pm}=\exp\left[K_{2}\sum_{m=1}^{M}(c_{m}^{\dag}-c_{m})(c_{m+1}^{\dag}+
c_{m+1})\right],
\end{equation}
and the $(+)$ sign in $V_2^+$ corresponds to the even states, to which are 
assigned the antiperiodic boundary conditions and the $(-)$ sign to the odd 
states, to which are assigned the periodic boundary conditions. 

Passing in a standard way to the momentum representation
\begin{eqnarray*}
c_{m}=\frac{\exp(-i\pi/4)}{\sqrt{M}}\sum_{q}e^{iq\*m}{\eta}_{q},
\end{eqnarray*}
we obtain after some simple transformations on $Z^*$, $(3.9)$ the expression:
\begin{equation}
Z^{*}=[2\cosh(h)]^{M}<0\mid(\prod_{0{\leq}q{\leq}{\pi}}^{}V_{2q}V_{hq}^{*})
\mid0>,
\end{equation}
where
\begin{eqnarray*}
V_{2q}=\exp\left\{2K_{2}\left[({\eta}_{q}^{\dag}{\eta}_{q}+
{\eta}_{-q}^{\dag}{\eta}_{-q}){\cos q}+({\eta}_{q}{\eta}_{-q}+
{\eta}_{-q}^{\dag}{\eta}_{q}^{\dag}){\sin q}\right]\right\}, \\
V_{hq}^{*}=\exp\left[{\alpha}^{2}\left({\frac{1+\cos q}{\sin q}}{\eta}_{-q}^{\dag}
{\eta}_{q}^{\dag}+f(q)+f(-q)\right)\right],
\end{eqnarray*}
in which the terms $f({\pm}q)$ in the expression for $V_{hq}^{*}$:
\begin{eqnarray*}
f(q)\equiv{-{\frac{1+e^{-iq}}{2\sin q}}{\eta}_{0}^{\dag}{\eta}_{q}^{\dag}}
\end{eqnarray*}
and in the case of antiperiodic boundary conditions should be omitted. 

Finally, calculating the vacuum matrix element for fixed $q$, after some not 
complicated transformations, we get for $Z^*$ $(3.11)$ in the case of even 
states the expression $(Z_+^*)$:
\begin{eqnarray}
Z_{+}^{*}=[2\cosh(h)]^{M}{\prod_{0<q<\pi}^{}}
[\cosh 2K_{2}-\sinh 2K_{2}\cos q+{\alpha}^{2}\sinh 2K_{2}(1+\cos q)]\nonumber\\
=[2{\cosh(h)}{\cosh K_{2}}]^{M}\prod_{m=1}^{M}\left[1+z_{2}^{2}+2z_{2}y-
2z_{2}(1-y)\cos[\frac{\pi(2m-1)}{M}]\right]^{1/2},
\end{eqnarray}
where $z_{2}\equiv{\tanh K_{2}}$ and  $y\equiv{\alpha}^{2}={\tanh^{2}h}$. 
Obviously, for $N$ noninteracting Ising models in external magnetic field the 
statistical sum $W(h)$ is equal to the $N-th$ power of the expression $(3.12)$, 
i.e. $W(h)=[Z_{+}^{*}]^{N}$. In the case of odd states, as one can easily 
show the following equality is satisfied:
\begin{equation}
Z_{-}^{*}=2Z_{+}^{*} .
\end{equation}
Let us note here that the representation $(3.12)$ unexpectedly finds an 
application in graph theory. Namely, with help of the representation $(3.12)$ 
one can calculate the generating function for Hamilton cycles on the simple 
rectangular lattice $(N\times M)$, $\cite{koch95}$. 

Finally, we obtain the following expression for free energy per spin in 
the thermodynamic limit:
\begin{equation}
-{\beta}F=\lim_{M\rightarrow\infty}\frac{1}{M}\ln Z^{*}=\ln{\left[e^{K_{2}}
\cosh h+(e^{2K_{2}}{\sinh^{2}h}+e^{-2K_{2}})^{1/2}\right]},
\end{equation}
i.e. the known classic expression $\cite{baxter82,izyum87}$. We paid so much 
attention to the $1D$ Ising model because we wanted to show on the first place 
to show effectiveness of the proposed method of transformation of the magnetic 
operator $V_h$ $(2.7)$ to its equivalent $(3.8)$. Additionally, as we mentioned 
above, a bit different representation of the statistical sum for the $1D$ Ising 
model $(3.11)$ finds its application in graph theory $\cite{koch95}$. Finally, 
this will help us to save time and place considerably when we will discuss the 
$2D$ and $3D$ Ising models in external magnetic field.

\section{TRANSFER-MATRIX}
In this section we will consider shortly the representation of the statistical 
sum for the $3D$ Ising model in external magnetic field $H$, applying for 
this purpose the well known transfer-martix method $\cite{baxter82,izyum87,huang63,thompson88}$. 
An exhaustive and outstanding presentation of the method the reader can find 
in the monographs $\cite{baxter82,izyum87}$, where are also shown other 
necessary pieces of knowledge on application of this method. 

Let us consider a simple cubic lattice consisting of $N$ rows, $M$ columns, 
and $K$ planes, in the sites of which the "spins" $\sigma_{nmk}$ are situated, 
which take on two values: $\sigma_{nmk}=\pm 1$. The Hamiltonian for the $3D$ 
Ising model in external magnetic field $H$ with the nearest-neighbour 
interaction is given in the form: 
\begin{equation}
{\cal H}=-\sum^{NMK}_{(n,m,k)=1}\left(J_{1}\sigma_{nmk}\sigma_{n+1,mk}+
J_{2}\sigma_{nmk}\sigma_{n,m+1,k}+J_{3}\sigma_{nmk}\sigma_{nm,k+1}+
H\sigma_{nmk}\right),
\end{equation}
where the multiindex $(nmk)$ numbers the sites of the simple cubic lattice 
$(N\times M\times K)$, and $H$ is the external magnetic field directed 
"upwards" $(\sigma_{nmk}=+1)$. The constants $(J_j>0)$ take into account 
anisotropy of the interaction of the Ising spins. There are periodic boundary 
conditions imposed, as usual, on the variables $\sigma_{nmk}$. The statistical 
sum for the system $Z_3(h)$ we write in the form:
\begin{eqnarray*}
Z_3(h)=\sum_{\sigma_{111}=\pm 1} ... \sum_{\sigma_{NMK}=\pm 1}e^{-\beta\cal H}=
\end{eqnarray*}
\begin{equation}
\sum_{\{\sigma_{nmk}=\pm 1\}}\exp\left[\sum_{nmk}(K_1\sigma_{nmk}
\sigma_{n+1,mk}+K_2\sigma_{nmk}\sigma_{n,m+1,k}+K_3\sigma_{nmk}
\sigma_{nm,k+1}+h\sigma_{nmk})\right],
\end{equation}
where the quantities $K_i$ and $h$ are defined as above $(2.3)$, [here and 
everywhere below summation over $nmk$ (or $nm$) and also multiplication over 
$nm$ will mean summation or multiplication over the full set of integer 
numbers from $1$ to $N,M$ and $K$ over each corresponding index, respectively]. 

In analogy to the two-dimensional case, it is convenient to introduce the 
notion of $k$-layer which is understood as a set of Ising spins in all sites 
of a $k$-layer:
\begin{eqnarray*}
a_{\{nm\}}\equiv a_{k}=\{\sigma_{nmk}\}, \;\;\;\; k-{\rm fixed} .
\end{eqnarray*}
Then summation in $(2.4)$ can be conveniently executed over the layers $a_k$, 
after writing the expression for $Z_3(h)$ in the form:
\begin{eqnarray*}
Z_3(h)=\sum_{a_1}\cdots\sum_{a_K}\exp\left\{\sum^{K}_{k=1}\left[\sum_{nm}
(K_1\sigma_{n+1,mk}+K_2\sigma_{n,m+1,k}+K_3\sigma_{nm,k+1}+h)\sigma_{nmk}
\right]\right\}
\end{eqnarray*}
\begin{equation}
=\sum_{a_1}\cdots\sum_{a_K}T^{\{\sigma_{nm1}\}}_{\{\sigma_{nm2}\}}
T^{\{\sigma_{nm2}\}}_{\{\sigma_{nm3}\}}\cdots T^{\{\sigma_{nmK}\}}_{\{\sigma
_{nm,K+1}\}},
\end{equation}
where
\begin{equation}
T^{\{\sigma_{nmk}\}}_{\{\sigma_{nm,k+1}\}}=\exp\left[\sum_{nm}(K_1\sigma_{
n+1,mk}+K_2\sigma_{n,m+1,k}+h)\sigma_{nmk}\right]\exp\left[K_3\sum_{nm}
\sigma_{nmk}\sigma_{nm,k+1}\right] .
\end{equation}
We will impose now periodic boundary conditions on the indices $n,m$, and $k$, 
taking 
\begin{equation}
\sigma_{N+1,mk}=\sigma_{1mk}, \;\;\;\; \sigma_{n,M+1,k}=\sigma_{n1k}, \;\;\;\;
\sigma_{nm,K+1}=\sigma_{nm1}
\end{equation}
As a consequence of what was stated above and of the conditions $(4.5)$ we 
can write $Z_3(h)$ in the form
\begin{equation}
Z_3(h)=Tr(T)^{K},
\end{equation}
where $T$ is the transfer-matrix,  matrix elements of which are described by 
equalities $(4.4)$. Matrix elements of the transfer-matrix of the layer-layer 
Ising model can be written in a bit different form $\cite{baxter82}$, but all 
these representations are in fact equivalent. Accordingly to the formula 
$(4.4)$ the matrix $T$ can be represented in the form of a product of the 
matrices $T_{1,2,3}$ and $T_h$, each of the same dimension $(2^{NM}\times 2^{NM})$: 
\begin{equation}
T=T_3T_2T_1T_h ,
\end{equation}
where
\begin{equation}
T_{3,b_{11}...b_{NM}}^{\phantom{3,}a_{11}...a_{NM}}=\prod_{nm}e^{K_{3}a_{nm}
b_{nm}},
\end{equation}
\begin{equation}
T_{2,b_{11}...b_{NM}}^{\phantom{2,}a_{11}...a_{NM}}=\delta_{a_{11}b_{11}}...\delta_{a_{NM}
b_{NM}}\prod_{nm}e^{K_{2}a_{nm}a_{n,m+1}},
\end{equation}
\begin{equation}
T_{1,b_{11}...b_{NM}}^{\phantom{1,}a_{11}...a_{NM}}=\delta_{a_{11}b_{11}}...\delta_{a_{NM}
b_{NM}}\prod_{nm}e^{K_{1}a_{nm}a_{n+1,m}},
\end{equation}
\begin{equation}
T_{h,b_{11}...b_{NM}}^{\phantom{h,}a_{11}...a_{NM}}=\delta_{a_{11}b_{11}}...\delta_{a_{NM}
b_{NM}}\prod_{nm}e^{ha_{nm}} .
\end{equation}
Here we introduced a new way of indexing the matrix elements in the expression 
$(4.4)$:
\begin{eqnarray*}
\{\sigma_{11k}\cdots\sigma_{NMk}\}\equiv\{a_{11}\cdots a_{NM}\}, \;\; 
\{\sigma_{11,k+1}\cdots\sigma_{NM,k+1}\}\equiv\{b_{11}\cdots b_{NM}\},
\end{eqnarray*}
and we will continue with these assignments till the end of the paper. 

Further as is known $\cite{huang63}$, if we introduce three sets of $2^{NM}$ - 
dimensional matrices $(\tau^{x,y,z}_{nm})$ of the form
\begin{equation}
\tau^{x,y,z}_{nm}=1\otimes 1\otimes\cdots\otimes\tau^{x,y,z}\otimes\cdots
\otimes 1\otimes 1, \;\;\; (NM -{\rm faktors}),
\end{equation}
where the Pauli matrices $\tau^{x,y,z}$ are situated in these products at the 
$nm$-th place, the matrices $T_{1,2,3}$ and $T_h$, $(4.8-11)$ can be rewritten 
in the form:
\begin{equation}
T_1=\exp\left(K_1\sum_{nm}\tau^{z}_{nm}\tau^{z}_{n+1,m}\right), \;\;\;\;
T_2=\exp\left(K_2\sum_{nm}\tau^{z}_{nm}\tau^{z}_{n,m+1}\right),
\end{equation}
\begin{equation}
T_3=(2\sinh 2K_3)^{NM/2}\exp\left(K^{*}_{3}\sum_{nm}\tau^{x}_{nm}\right),
\end{equation}
\begin{equation}
T_h=\exp\left(h\sum_{nm}\tau^{z}_{nm}\right),
\end{equation}
where the quantities $K_3$ and $K^*_3$ are connected by the conditions of the 
form $(2.8)$, and the spin Pauli matrices $\tau^{x,y,z}_{nm}$, $(4.12)$ 
commute one with each other for different $(nm)\neq (n'm')$, and simultaneously 
for each given $nm$ these matrices satisfy the standard conditions $\cite{huang63}$. 
It is easy to see that the matrices $T_{1,2,h}$, $(4.13,15)$ commute one with 
each other, but they do not commute with the matrix $T_3$, $(4.14)$. Obviously, 
for $(h=0)$ we obtain the known formulas $\cite{baxter82}$ for the matrices 
$T_{1,2,3}$, describing the three-dimensional Ising model on a simple cubic 
lattice. To the transition to the $2D$ Ising model in the interaction constants 
$K_1$ and $K_2$ corresponds taking $(K_1=0)$ or $(K_2=0)$ and simultaneously 
removal of summation over $n$ $(N=1)$ or over $m$ $(M=1)$ respectively. We 
obtain this way the standard expressions $\cite{sml64,izyum87}$ for the $2D$ 
Ising model in external magnetic field, and the operator $T_1$ $(4.13)$ is 
identically equal to the unit operator $(T_1\equiv 1)$ in the first case, and 
$(T_2\equiv 1)$ in the second case, respectively.

A bit different situation occurs in the case of transition to the $2D$ Ising 
model in the interaction constant $K_3$. In this case one should take 
$(K_3=0,  K=1)$, i.e. omit summation over $k$. As a result one can arrive at 
the following formula for the operator $T_3$ $(4.14)$:
\begin{equation}
T^{*}_{3}\equiv T_3(K_3=0)=\prod_{nm}(1+\tau^{x}_{nm}) .
\end{equation}
Namely this structure of the operator $T_3^*$ enables, finally, effective 
rebuilding of the magnetic operator $T_h$ $(4.15)$, as it was shown above on 
the example of the $1D$ Ising model. In this case we can write the expression 
for the statistical sum for the $2D$ Ising model in the form:
\begin{equation}
Z_2(h)=Tr(T^{*}_{3}T_2T_1T_h) ,
\end{equation}
where the matrices $T_{1,2,h}$ are defined by the formulas $(4.13,15)$, and 
the matrix $T_3^*$ is defined by the formula $(4.16)$. The advantage of the 
representation of the statistical sum $(4.17)$ is, in the opinion of the 
author, in a sense obviously. We will write about this issue additionally 
below. As it will be clear from what is stated further the matrix $T_2T_1T_h$ 
can be conveniently written in the form $T^{1/2}_hT_2T_1T^{1/2}_h$ , where 
we applied commutativity of the factors, following from commutativity of the 
matrices $\tau^z_{nm}$. The statistical sum $(4.17)$ we rewrite in the form:
\begin{equation}
Z_{2}(h)=Tr(T^{*}_{3}T^{1/2}_hT_2T_1T^{1/2}_h) ,
\end{equation}
where the matrix $T^{1/2}_h$ is defined by the formula
\begin{equation}
T_{h/2}\equiv T^{1/2}_h=\exp\left[(h/2)\sum_{nm}\tau^{z}_{nm}\right] .
\end{equation}
Below we will use both the expression $(4.17)$ and the representation $(4.18)$, 
having in mind further applications in graph theory $\cite{koch95,graph67}$.

\section{TRANSFORMATION OF $T$-OPERATOR}
\subsection{Introduction of Fermion operators} 

Schultz, Mattis and Lieb $\cite{sml64}$ showed that the $T$-matrix in its 
standard representation can be expressed in terms of the second quantization 
Fermi operators. For this aim they applied the known Jordan-Wigner transformations 
$\cite{jordan28}$ which enable expression of the Fermi operators $(c^{\dag}_{m},c_{m})$ 
for the one-dimensional system by the Pauli operators 
$(\tau^{\pm}_{m})$, $\cite{izyum87}$:
\begin{equation}
c_m=\exp{\left( i\pi\sum^{m-1}_{j=1}\tau^+_j\tau^-_j\right)}\tau^-_m, \;\;\;
c_m^{\dag}=\exp{\left( i\pi\sum^{m-1}_{j=1}\tau^+_j\tau^-_j\right)}\tau^+_m .
\end{equation}
As was shown in $\cite{mkoch95}$, there is an analogue to the Jordan-Wigner 
transformations $(5.1)$ which generalizes the former to the two-, three-, and 
$d$-dimensional systems. 

For this aim we introduce first the following variables $\cite{izyum87}$ 
to the formulas $(4.13-16)$:
\begin{equation}
\tau^{\pm}_{nm}=\frac{1}{2}(\tau^{z}_{nm}\pm i\tau^{y}_{nm}),
\end{equation}
which satisfy anticommutation relations for the same site:
\begin{equation}
\{\tau^{+}_{nm},\tau^{-}_{nm}\}_{+}=1, \;\;\;\;\;\; (\tau^{+}_{nm})^{2}=
(\tau^{-}_{nm})^{2},
\end{equation}
and commutation relations for various sites, 
\begin{equation}
[\tau^{\pm}_{nm},\tau^{\pm}_{n'm'}]_{-}=0, \;\;\;\;\;\; (nm)\neq (n'm').
\end{equation}
Quantities $\tau^{\pm}_{nm}$ are often called Pauli operators. The correspondences 
\begin{equation}
\tau^{x}_{nm}=-2(\tau^{+}_{nm}\tau^{-}_{nm}-1/2), \;\;\;\;\;\;\; \tau^{z}_{nm}=
\tau^{+}_{nm}+\tau^{-}_{nm},
\end{equation}
enable to rewrite the expressions for $T_{1,2,h}$ and $T_3^*$, $(4.13-16)$ in 
the form:
\begin{equation}
T_1=\exp\left[K_1\sum_{nm}(\tau^{+}_{nm}+\tau^{-}_{nm})(\tau^{+}_{n+1,m}+
\tau^{-}_{n+1,m})\right],
\end{equation}
\begin{equation}
T_2=\exp\left[K_2\sum_{nm}(\tau^{+}_{nm}+\tau^{-}_{nm})(\tau^{+}_{n,m+1}+
\tau^{-}_{n,m+1})\right],
\end{equation}
\begin{equation}
T_h=\exp\left[h\sum_{nm}(\tau^{+}_{nm}+\tau^{-}_{nm})\right],
\end{equation}
\begin{equation}
T^{*}_{3}=\prod_{nm}\left[1+(1-2\tau^{+}_{nm}\tau^{-}_{nm})\right].
\end{equation}

As it was mentioned above, the Pauli operators $\tau^{\pm}_{nm}$ behave as 
Fermi operators when considered for one site, and as Bose operators when 
considered for different sites. In order to transform to the fermionic  
representation, i.e. to the Fermi operators in the whole lattice, we will 
introduce an analogue of the Jordan-Wigner transformations $(5.1)$, which will 
enable to express Fermi operators $(c^{\dag}_{nm},c_{nm})$ by Pauli operators 
$\tau^{\pm}_{nm}$ for the two-dimensional system. Namely, there exist in the 
two-dimensional case two sets of such transformations $\cite{mkoch95}$, which 
we represent here in the form:
\begin{eqnarray*}
\alpha^{\dag}_{nm}=\exp\left( i\pi\sum^{n-1}_{k=1}\sum^{M}_{l=1}\tau^+_{kl} 
\tau^-_{kl}+i\pi\sum^{m-1}_{l=1}\tau^+_{nl}\tau^-_{nl}\right)\tau^+_{nm},
\end{eqnarray*}
\begin{equation}
\alpha_{nm}=\exp\left( i\pi\sum^{n-1}_{k=1}\sum^{M}_{l=1}\tau^+_{kl} 
\tau^-_{kl}+i\pi\sum^{m-1}_{l=1}\tau^+_{nl}\tau^-_{nl}\right)\tau^-_{nm},
\end{equation}
and
\begin{eqnarray*}
\beta^{\dag}_{nm}=\exp\left( i\pi\sum^{N}_{k=1}\sum^{m-1}_{l=1}\tau^+_{kl} 
\tau^-_{kl}+i\pi\sum^{n-1}_{k=1}\tau^+_{km}\tau^-_{km}\right)\tau^+_{nm},
\end{eqnarray*}
\begin{equation}
\beta_{nm}=\exp\left( i\pi\sum^{N}_{k=1}\sum^{m-1}_{l=1}\tau^+_
{kl} \tau^-_{kl}+i\pi\sum^{n-1}_{k=1}\tau^+_{km}\tau^-_{km}\right)\tau^-_{nm}.
\end{equation}
It is easy to show, using formulas $(5.3-4)$ that the operators 
$(\alpha^{\dag}_{nm},\alpha_{nm})$ and $(\beta^{\dag}_{nm},\beta_{nm})$ are Fermi 
operators in the whole lattice, i.e. they satisfy anticommutation relations 
for all sites:
\begin{equation}
\{\alpha^{\dag}_{nm},\alpha_{nm}\}_{+}=1,\;\; (\alpha^{\dag}_{nm})^2=(\alpha_{nm})^
2=0;\;\; 
\{\alpha^{\dag}_{nm},\alpha^{\dag}_{n'm'}\}_{+}= ...=0,\;\; (nm)\neq (n'm'),
\end{equation}
and analogously for the $\beta$-operators. There are also inverse transformations: 
\begin{equation}
\tau^{+}_{nm}=\exp\left[i\pi\sum^{n-1}_{k=1}\sum^{M}_{p=1}\alpha^{\dag}_{kp}
\alpha_{kp}+i\pi\sum^{m-1}_{p=1}\alpha^{\dag}_{np}\alpha_{np}\right]\alpha^{\dag}_{nm},
\end{equation}
etc., which can be easily proved by application of the identities
\begin{eqnarray*}
\exp\left(i\pi\sum_{nm}\tau^{+}_{nm}\tau_{nm}\right)=\prod_{nm}\left(1-
2\tau^{+}_{nm}\tau^{-}_{nm}\right)=\prod_{nm}\tau^{x}_{nm},
\end{eqnarray*}
from which one can easily derive the equalities 
\begin{equation}
\tau^{+}_{nm}\tau^{-}_{nm}=\alpha^{\dag}_{nm}\alpha_{nm}=\beta^{\dag}_{nm}\beta_{nm}.
\end{equation}
The formulas $(5.14)$ express conditions of local equality of the occupation 
numbers for $\alpha$- and $\beta$- fermions in one site. Further, as it follows 
from $(5.10-11)$ and $(5.13)$, $\alpha$- and $\beta$- operators are connected 
by canonical non-linear transformations:
\begin{eqnarray*}
\alpha^{\dag}_{nm}=\exp(i\pi\varphi_{nm})\beta^{\dag}_{nm}, \;\;\;\; 
\alpha_{nm}=\exp(i\pi\varphi_{nm})\beta_{nm},
\end{eqnarray*}
\begin{equation}
\varphi_{nm}=\left[\sum^{N}_{k=n+1}\sum^{m-1}_{p=1}+\sum^{n-1}_{k=1}
\sum^{M}_{p=m+1}\right]\alpha^{\dag}_{kp}\alpha_{kp}=[\cdots]\beta^{\dag}_{kp}
\beta_{kp},
\end{equation}
where the operators $\varphi_{nm}$ obviously commute with the operators 
$(\alpha^{\dag}_{nm},\alpha_{nm})$ and $(\beta^{\dag}_{nm},\beta_{nm})$, i.e. 
\begin{equation}
[\varphi_{nm},\alpha^{\dag}_{nm}]_{-}=\cdots=\cdots=[\varphi_{nm},\beta_{nm}]_{-}
=0 .
\end{equation}
Commutation relations among $\alpha$- and $\beta$- operators are more 
complicated. Namely, as one can check by direct calculation that the following 
commutation relations hold:
\begin{equation}
\{\alpha^{\dag}_{nm},\beta_{nm}\}_{+}=\{\beta^{\dag}_{nm},\alpha_{nm}\}_{+}
=(-1)^{\varphi_{nm}},
\end{equation}
\begin{equation}
\left[\alpha_{nm},\beta_{n'm'}\right]_{-}=\ldots=[\alpha^{\dag}_{nm},
\beta^{\dag}_{n'm'}]_-=0,\;\;\;\;\;
\left(\begin{array}{c c}n'\leq n-1,&m'\geq m+1\\
n'\geq n+1,&m'\leq m-1  \end{array}\right),
\end{equation}
and
\begin{equation}
\{\alpha_{nm},\beta_{n'm'}\}_+ = \ldots=\{\alpha^{\dag}_{nm},\beta^{\dag}_{n'm'}\}_+=0,
\end{equation}
in all other cases, where $\varphi_{nm}$ are defined above $(5.15)$. This way 
we get rather specific structure of commutation relations among $\alpha$- and 
$\beta$- operators in the lattice, although this structure shows some symmetry. 
Here is the right place to compare the situation described above with the 
situation we get using the second quantization method. For a system composed 
of different particle one introduces the second quantization operators of 
different kinds for different particles. The operators connected to either 
bosons or fermions satisfy the standard commutation relations. As far as the 
operators for different fermions are concerned, it is usually assumed without 
any proof $\cite{landau89}$, that within the limits of nonrelativistic theory 
they could be treated formally as commuting or anticommuting. Both assumptions 
lead to the same results when the second quantization method is applied. 
Nevertheless, in the relativistic theory, which allows for transmutations of 
various particles we should consider creation and annihilation operators for 
different fermions as anticommuting. On the other hand, in our case we deal 
formally with "quasiparticles" of the $\alpha$- and $\beta$- types underlying 
separately the Fermi statistics with commutation relations among particles of 
different types being however dependent on relative position of these 
"quasiparticles" in the sites of lattice. Such a situation, as far as is 
known to the author, was not present in earlier works on application of the 
second quantization method.

\subsection{The $T_{1,2,h}$- and  $T^{*}_{3}$ - operators}

Before writing the $T$- operators $(5.6-9)$ in terms of Fermi operators, let 
as make a few remarks. First, the operator $T_3^*$ $(5.9)$ can be expressed 
in terms of $\alpha$ as well as $\beta$-operators, because of $(5.14)$:
\begin{equation}
T^{*}_3=\prod_{nm}\left[1+(-1)^{\alpha^{\dag}_{nm}\alpha_{nm}}\right]=
\prod_{nm}\left[1+(-1)^{\beta^{\dag}_{nm}\beta_{nm}}\right],
\end{equation}
where the basic in the Fock representation should be chosen as to be expressed 
in terms of the $\alpha$- or $\beta$- operators, respectively. Second, the 
operators $T_{1,2,h}$ we can also express in terms of either $\alpha$- or 
$\beta$- operators. Nevertheless, the operator $T_2$ we write in terms of the 
$\alpha$- operators and the operator $T_1$ we write in terms of the $\beta$- 
operators for reasons which will become clear later. 

Now, due to $(.10-11)$ we can write the operator $T_h$ $(5.8)$ in the form:
\begin{equation}
T_h=\exp\left[h\sum_{nm}\theta_{nm}(\alpha^{\dag}_{nm}+\alpha_{nm})\right]=
\exp\left[h\sum_{nm}\psi_{nm}(\beta^{\dag}_{nm}+\beta_{nm})\right],
\end{equation}
where $\theta_{nm}$ is defined as the first factor in $(5.13)$, and 
$\psi_{nm}$ is defined by:
\begin{eqnarray*}
\psi_{nm}=\exp\left[i\pi\sum^{N}_{k=1}\sum^{m-1}_{p=1}\beta^{\dag}_{kp}\beta_{kp}
+i\pi\sum^{n-1}_{k=1}\beta^{\dag}_{km}\beta_{km}\right].
\end{eqnarray*}

Transformation of the operators $T_{1,2}$ is a bit more complicated. Taking 
into account cyclic boundary conditions $(4.5)$, we will write first a sequence 
of equalities analogous to $(5.14)$:
\begin{eqnarray*}
\tau^{+}_{N,m}\tau^{+}_{1,m}=-(-1)^{{\hat N}_m}\beta^{\dag}_{N,m}\beta_{1,m}^
{\dag},\;\;\;\;\;
\tau^{+}_{N,m}\tau^{-}_{1,m}=-(-1)^{{\hat N}_m}\beta^{\dag}_{N,m}\beta_{1,m},
\end{eqnarray*}
\begin{equation}
\tau^{-}_{N,m}\tau^{+}_{1,m}=(-1)^{{\hat N}_m}\beta_{N,m}\beta_{1,m}^{\dag}, \;\;\;\;\;
\tau^{-}_{N,m}\tau^{-}_{1,m}=(-1)^{{\hat N}_m}\beta_{N,m}\beta_{1,m},
\end{equation}
and
\begin{eqnarray*}
\tau^{+}_{n,M}\tau^{+}_{n,1}=-(-1)^{{\hat M}_n}\alpha^{\dag}_{n,M}\alpha^
{\dag}_{n,1}, \;\;\;\;\;
\tau^{+}_{n,M}\tau^{-}_{n,1}=-(-1)^{{\hat M}_n}\alpha^{\dag}_{n,M}\alpha_{n,1},
\end{eqnarray*}
\begin{equation}
\tau^{-}_{n,M}\tau^{+}_{n,1}=(-1)^{{\hat M}_n}\alpha_{n,M}\alpha^{\dag}_{n,1}, \;\;\;\;\;
\tau^{-}_{n,M}\tau^{-}_{n,1}=(-1)^{{\hat M}_n}\alpha_{n,M}\alpha_{n,1},
\end{equation}
where
\begin{equation}
{\hat g}_n\equiv(-1)^{{\hat M}_n},\;\;\; {\hat M}_n=\sum^{M}_{m=1}\alpha^{\dag}_{nm}
\alpha_{nm}; \;\;\;\;\;{\hat g}_m\equiv(-1)^{{\hat N}_m},\;\;\; {\hat N}_m=\sum^{N}_
{n=1}\beta^{\dag}_{nm}\beta_{nm},
\end{equation}
which can be obtained by using the formulas $(5.10-13)$. Therefore we can 
write the following representations for the operators $T_{1,2}$:
\begin{equation}
T_1=\exp\left\{K_1\sum^{M}_{m=1}\left[\sum^{N-1}_{n=1}(\beta^{\dag}_{nm}-
\beta_{nm})(\beta^{\dag}_{n+1,m}+\beta_{n+1,m})-{\hat g}_m(\beta^{\dag}_{Nm}-
\beta_{Nm})(\beta^{\dag}_{1,m}+\beta_{1,m})\right]\right\},
\end{equation}
\begin{equation}
T_2=\exp\left\{K_2\sum^{N}_{n=1}\left[\sum^{M-1}_{m=1}(\alpha^{\dag}_{nm}-
\alpha_{nm})(\alpha^{\dag}_{n,m+1}+\alpha_{n,m+1})-{\hat g}_n(\alpha^{\dag}_{nM}-
\alpha_{nM})(\alpha^{\dag}_{n,1}+\alpha_{n,1})\right]\right\}.
\end{equation}
Finally, let us express the operator $T_2$ in terms of the $\beta$ -operators:
\begin{equation}
T_2\!=\!\exp\!\!\left\{\!K_2\!\sum^{N}_{n=1}\!\!\left[\sum^{M-1}_{m=1}{\hat{\chi}}_{nm}
(\beta^{\dag}_{nm}-\beta_{nm})(\beta^{\dag}_{n,m+1}+\beta_{n,m+1})-{\hat G}
{\hat{\chi}}_{nM}(\beta^{\dag}_{nM}-\beta_{nM})(\beta^{\dag}_{n,1}+\beta_{n,1})
\right]\!\!\right\},
\end{equation}
where the operators ${\hat G}$ and ${\hat{\chi}}_{nm}$ are defined by the formulas: 
\begin{eqnarray*}
{\hat G}\equiv\exp\left[i\pi\sum_{nm}\alpha^{\dag}_{nm}\alpha_{nm}\right]=
\exp\left[i\pi\sum_{nm}\beta^{\dag}_{nm}\beta_{nm}\right]=(-1)^{{\hat S}},
\end{eqnarray*}
\begin{equation}
{\hat{\chi}}_{nm}=\exp\left[i\pi\sum^{N}_{k=n+1}\beta^{\dag}_{km}\beta_{km}+
i\pi\sum^{n-1}_{k=1}\beta^{\dag}_{k,m+1}\beta_{k,m+1}\right],
\end{equation}
and we applied the relations analogous to $(5.23)$, but expressed in terms of 
the $\beta$- operators. The operator ${\hat S}$ introduced above $(5.28)$ is 
the operator of the number of particles, which is connected with the operators 
${\hat N}$ and ${\hat M}$, $(5.24)$ by relations:
\begin{equation}
{\hat S}=\sum^{N}_{n=1}{\hat M}_n=\sum^{M}_{m=1}{\hat N}_m, \;\;\;\;\;\;\;
{\hat G}=\prod^{N}_{n=1}{\hat g}_n=\prod^{M}_{m=1}{\hat g}_m .
\end{equation}
It is easy to see that the operator ${\hat G}$, $(5.28)$ commutes with the 
operators $T_1$ and $T_2$ $(5.25-27)$, but it does not commute with the 
operator $T_h$ $(5.21)$, because the following relations are satisfied: 
\begin{equation}
\{{\hat G},\alpha^{\dag}_{nm}\}_+=\cdots =\{{\hat G},\beta_{nm}\}_+=0
\end{equation}
Of course, we can also express the operators $T_1$ and $T_h$ in terms of the 
$\alpha$- operators and we can write down the formulas if they are necessary. 

It was shown above $(4.17)$ that the statistical sum for the $2D$ Ising model 
in external magnetic field can be represented by the trace of the operator 
$T$, which was expressed here by the Fermi second quantization operators. 
Introducing, as in the one-dimensional case, a basic in the occupation 
numbers representation $\cite{landau89}$ for the $\alpha$- and $\beta$- 
fermions ($2^{NM}$ - dimensional space in the Fock representation), and 
calculating then appropriate matrix elements $<l|T|l>$, it is easy to see 
that because of multiplicative character of the operator $T^*_3$, $(5.20)$ all 
matrix elements, besides the vacuum matrix element $<0|T|0>$, are equal to 
zero. For the vacuum matrix element contribution from the operator $T^*_3$ is 
equal simply to $2^{NM}$, and we can write $Z_2(h)$ $(4.17-18)$ in the 
following form:
\begin{equation}
Z_2(h)=2^{NM}<0|(T_2T_1T_h)|0>=2^{NM}<0|(T_{h/2}T_2T_1T_{h/2})|0>,
\end{equation}
where the vacuum state $|0>$ is defined in the standard manner
\begin{equation}
\alpha_{nm}|0>=\beta_{nm}|0>=0, \;\;\;\;\;n(m)=1,2, ..., N(M),
\end{equation}
and operators $T_{1,2,h}$ are defined by the formulas $(5.21)$ and $(5.25-27)$. 
Let us stress that the vacuum state $(5.32)$ for the $\alpha$- and $\beta$- 
fermions can differ among themselves at most by a constant phase factor, which 
in the given case can always be taken to be equal to unity. However, it is no 
longer true in the case of multiparticle states for the $\alpha$- and $\beta$- 
fermions, because in this case the essential role begin to play phase factors 
$(-1)^{\varphi_{nm}}$, $(5.15)$. As an exception serve the one-particle states, 
for which, as it can be easily found from $(5.15)$, we have:
\begin{eqnarray*}
\alpha^{\dag}_{nm}|0>=(-1)^{\varphi_{nm}}\beta^{\dag}_{nm}|0>=\beta^{\dag}_{nm}|0> ,
\end{eqnarray*}
for all$(nm)$. In all other cases the $\alpha$- and $\beta$- states will differ 
one from each other by their sign which depends on indices $(nm)$ of the 
corresponding sites. This very fact implies main difficulty in the proposed 
approach to solving the problem under consideration. This difficulty can be, 
however, avoided.

Let us make two remarks here. It is obvious that the representation $(55.31)$ 
for the statistical sum $Z_2(h)$ does not depend on the kind of variables 
($\alpha$- or $\beta$- operators) with which we introduce the basic in the 
representation of occupation numbers, because equality of local occupation 
numbers $(5.14)$ holds for the $\alpha$- and $\beta$- fermions. Further, we 
expressed the operator $T_2$ in terms of the $\alpha$- and $\beta$- variables 
$(5.26-27)$, although we will work mainly with the expression $(5.26)$. The 
reason is that in the representation $(5.27)$ for $T_2$ the operators 
${\hat{\chi}}_{nm}$ are present $(5.28)$. They are phase factors and it is 
difficult in practice to remove them. Difficulty coming from the presence of 
these operators is of the same kind, which was found by the authors of the 
paper $\cite{sml64}$  who considered the case with external magnetic field. 
Simultaneously, the representation $(5.26)$ for $T_2$ does not involve the 
phase factors, justifying the choice. Nevertheless, the expression $(5.27)$ 
for $T_2$ will be necessary in the analysis of the boundary conditions, which 
play here important role. Analogous statement applies to the operator $T_1$, 
which we expressed in terms of the $\alpha$- and $\beta$- variables $(5.25)$ 
and which also does not contain phase factors of the type of ${\hat{\chi}}_{nm}$. 
The essence of our approach lies in the structure of the transformations 
$(5.10-11)$, which allows for expression of the operators $T_{1,2}$ in the 
form $(5.26)$, which does not contain the phase factors. 

Now, we transform the magnetic operator $T_h$ $(5.21)$, or more exactly the 
ket-vector $T_h|0>$, entering the expression $(5.31)$ for $Z_2(h)$. Exactly 
in the sense one should understand the equivalence of the two operators $T_h$ 
and $T_h^*$, acting on the vacuum state $|0>$. Below we will omit $|0>$, as 
this should not lead to misunderstandings. Analogously, we introduce the 
notation $T^{l,r}_{h/2}$ for the transformed bra-vector $<0|T_{h/2}$ and the 
transformed ket-vector $T_{h/2}|0>$, respectively, omitting further bra- and 
ket-vectors of the vacuum state $(<0|, |0>)$. Continuing with considerations 
analogous to these, which gave us the expression $(3.7)$ in the one-dimensional 
case, the operator $T_h$ $(5.21)$ we represent in the form:
\begin{equation}
T_h\!=\!(\cosh h)^{NM}\!\exp\!\left[\alpha\sum_{nm}\beta^{\dag}_{nm}\right]
\!\exp\!\left\{\!\alpha^2\left[\sum^{N}_{n,n'}\sum^{M}_{m=1}\sum^{M-m}_{p=1}
\beta^{\dag}_{nm}\beta^{\dag}_{n',m+p}+\sum^{N}_{n=1}\sum^{N-n}_{k=1}\sum^{M}_{m=1}
\beta^{\dag}_{nm}\beta^{\dag}_{n+k,m}\right]\right\},
\end{equation}
where $\alpha\equiv\tanh h$. Analogously, the operators $T^{l,r}_{h/2}$ we 
write in the form:
\begin{equation}
T_{h/2}^l\!=\!(\cosh\frac{h}{2})^{NM}\!\exp\!\left[\mu\!\sum_{nm}\alpha_{nm}\right]
\!\exp\!\left\{\!\mu^2\!\!\!\left[\sum^{N}_{n=1}\sum^{M}_{m=1}\sum^{M-m}_{p=1}
\alpha_{n,m+p}\alpha_{nm}+\sum^{N}_{n=1}\sum^{N-n}_{k=1}\sum^{M}_{m,m'}
\alpha_{n+k,m'}\alpha_{nm}\right]\!\right\},
\end{equation}
\begin{equation}
T_{h/2}^r\!=\!(\cosh\frac{h}{2})^{NM}\!\exp\!\left[\mu\!\sum_{nm}\beta^{\dag}_{nm}\right]
\!\exp\!\left\{\!\mu^2\!\!\!\left[\sum^{N}_{n,n'}\sum^{M}_{m=1}\sum^{M-m}_{p=1}
\beta^{\dag}_{nm}\beta^{\dag}_{n',m+p}+\sum^{N}_{n=1}\sum^{N-n}_{k=1}\sum^{M}_{m=1}
\beta^{\dag}_{nm}\beta^{\dag}_{n+k,m}\right]\!\right\},
\end{equation}
where $\mu\equiv\tanh(h/2)$. The operators $T_h$ and $T^{l,r}_{h/2}$ are of 
rather complicated structure. However, they do not contain the phase factors 
any longer. Substituting the expressions $(5.33-35)$ to the equalities $(5.31)$, 
the statistical sum $Z_2(h)$ can be written in the form: 
\begin{equation}
Z_2(h)=2^{NM}<0|(T_2T_1T^{*}_h)|0>,
\end{equation}
or
\begin{equation}
Z_2(h)=2^{NM}<0|(T^{*}_lT_2T_1T^{*}_r+\mu^{2}AT^{*}_lT_2T_1T^{*}_rB)|0>\equiv
2^{NM}<0|(U_1+U_2)|0>,
\end{equation}
where the operators $U_{1,2}$ are defined in the obvious way, and the operators 
$T^{*}_h$ and $T^{*}_{l,r}$ are given by the formulas $(5.33-35)$, in which 
one should omit the factors
\begin{eqnarray*}
\exp\left(\alpha\sum_{nm}\beta^{\dag}_{nm}\right), \;\;\;\;
\exp\left(\mu\sum_{nm}\alpha_{nm}\right), \;\;\;\;
\exp\left(\mu\sum_{nm}\beta^{\dag}_{nm}\right),
\end{eqnarray*}
and the operators $A$ and $B$ are of the form:
\begin{equation}
A=\sum_{nm}\alpha_{nm}, \;\;\;\;\;\;\; B=\sum_{nm}\beta^{\dag}_{nm}.
\end{equation}
In derivation of $(5.36-37)$ we used the fact that the diagonal matrix elements 
for the product of odd number of Fermi operators is equal to zero, and that 
the following equalities are true 
\begin{eqnarray*}
\exp\left[a\sum_{nm}\alpha_{nm}(\beta^{\dag}_{nm})\right]=
1+a\sum_{nm}\alpha_{nm}(\beta^{\dag}_{nm}),
\end{eqnarray*}
where $a$ is a $c$-valued function. One should remember also that the operators 
$T_{1,2}$ $(5.25-27)$ contain only bilinear products of the Fermi operators. 
With this ends the rebuilding procedure for the magnetic operator. 

\subsection{The Boundary Conditions} 

With the aim of further simplification of the operators $T_{1,2}$ we should 
consider boundary conditions for the $\alpha$- and $\beta$- operators, taking 
periodic boundary conditions for the Pauli operators $\tau^{\pm}_{nm}$ $(5.2)$ 
in both indices $(nm)$ as a starting point. Let us shortly discuss this problem 
here. First, since all terms in $T_{1,2}$ and $T^*_h$ contain bilinear products 
of the Fermi operators, the following formulas are valid:
\begin{equation}
[{\hat G},\;T_1]_{-}=[{\hat G},\;T_2]_{-}=[{\hat G},\;T^{*}_h]_{-}=\;0,
\end{equation}
which shows that the states with even or odd number of fermions are preserved 
as well for the $\alpha$- as for the $\beta$- particles. The operator ${\hat G}$, 
entering $(5.39)$, is defined above $(5.28)$. Analogously, the following 
formulas are true:
\begin{eqnarray*}
[{\hat G},\;U_1]_{-}\;=\;[{\hat G},\;U_2]_{-}=\;0,
\end{eqnarray*}
where the operators $U_{1,2}$ are defined above. However, the operators 
${\hat g}_{n}$ and ${\hat g}_m$ $(5.24)$ do not commute with the operators 
$T=T_2T_1T^{*}_h$ or $U=U_1+U_2$ $(5.37)$, and this fact implies some difficulties. 
First, let us note that the following equalities hold:
\begin{equation}
{\hat G}=\prod_{n}{\hat g}_{n}=\prod_{m}{\hat g}_{m},\;\;\;\;\;\;
\lambda_{\hat G}=\prod_{n}\lambda_{{\hat g}_n}=\prod_{m}\lambda_{{\hat g}_m},
\end{equation}
where by $\lambda_{\hat G}$, $\lambda_{{\hat g}_n}$ and $\lambda_{{\hat g}_m}$ 
we denoted eigenvalues of the operators ${\hat G}$, ${\hat g}_n$ and 
${\hat g}_m$, equal ${\pm 1}$. 

Let us consider first the case corresponding to the state with even number of 
fermions $(\lambda_{\hat G}=+1)$. In this case, as one can easily see, we 
should choose antiperiodic boundary conditions for the $\beta$- operators 
with respect to the second index $m$ (for all $n$), and antiperiodic boundary 
conditions for the $\alpha$- operators with respect to the first index $n$ 
(for all $m$), i.e. 
\begin{eqnarray*}
\beta^{\dag}_{n,M+1}=-\beta^{\dag}_{n,1}, \;\;\;\;\; \beta_{n,M+1}=-\beta_
{n,1},\;\;\;\; (n=1,2,...,N);
\end{eqnarray*}
\begin{equation}
\alpha^{\dag}_{N+1,m}=-\alpha^{\dag}_{1,m}, \;\;\;\;\; \alpha_{N+1,m}=-\alpha_{1,m},
\;\;\;\; (m=1,2,...,M).
\end{equation}
Then the boundary conditions for the $\beta$- operators with respect to the 
first index $n$ depend on ${\hat g}_m$, and the boundary conditions for the 
$\alpha$- operators with respect to the second index $m$ depend on ${\hat g}_n$. 
More exactly, it depends on at which step we fix the eigenvalues 
$\lambda_{{\hat g}_m}$ and $\lambda_{{\hat g}_n}$, respectively. The only 
limitations on the choice of the eigenvalues and corresponding boundary 
conditions give equalities $(5.40)$. The whole freedom in the choice of 
boundary conditions consist of $2^N$ possible boundary conditions for the 
$\alpha$- operators in their second index, and $2^M$ possible boundary 
conditions for the $\beta$- operators in their first index. Detailed analysis 
shows that we can without loosing generality choose homogeneous boundary 
conditions, i.e. antiperiodic boundary conditions for the $\alpha$- operators 
in their second index $m$, which corresponds to $(\lambda_{{\hat g}_n}=+1)$ for each $n$, 
and antiperiodic boundary conditions for the $\beta$- operators in their first 
index $n$, which corresponds to $(\lambda_{{\hat g}_m}=+1)$ for each $m$, i.e. 
\begin{eqnarray*}
\alpha^{\dag}_{n,M+1}=-\alpha^{\dag}_{n,1}, \;\;\;\; \alpha_{n,M+1}=-\alpha_{n,1}, 
\;\;\;\;\lambda_{{\hat g}_n}=+1,\;\;(n=1,2,....N);
\end{eqnarray*}
\begin{eqnarray*}
\beta^{\dag}_{N+1,m}=-\beta^{\dag}_{1,m},\;\;\;\; \beta_{N+1,m}=-\beta_{1,m}, 
\;\;\;\; \lambda_{{\hat g}_m}=+1,\;\; (m=1,2,...,M);
\end{eqnarray*}
\begin{equation}
\lambda_{\hat G}=\prod_{n}\lambda_{{\hat g}_n}=(+1)^N=\prod_{m}\lambda_{{\hat
g}_m}=(+1)^M=+1,
\end{equation}
for each parity of the numbers $N$ and $M$. Analogously, one can show that in 
the case of the odd states $(\lambda_{\hat G}=-1)$ the boundary conditions 
for the $\alpha$- and $\beta$- operators can be written in the form:
\begin{eqnarray*}
\alpha^{\dag}_{N+1,m}=+\alpha^{\dag}_{1,m}, \;\;\;\;\; \alpha_{N+1,m}=+\alpha_{1,m},
\;\;\; (m=1,2,...,M);
\end{eqnarray*}
\begin{eqnarray*}
\alpha^{\dag}_{n,M+1}=+\alpha^{\dag}_{n,1}, \;\;\;\; \alpha_{n,M+1}=+\alpha_{n,1}, 
\;\;\;\;\lambda_{{\hat g}_n}=-1,\;\;(n=1,2,....N);
\end{eqnarray*}
\begin{eqnarray*}
\beta^{\dag}_{N+1,m}=+\beta^{\dag}_{1,m},\;\;\;\; \beta_{N+1,m}=+\beta_{1,m}, 
\;\;\;\; \lambda_{{\hat g}_m}=-1,\;\; (m=1,2,...,M);
\end{eqnarray*}
\begin{eqnarray*}
\beta^{\dag}_{n,M+1}=+\beta^{\dag}_{n,1}, \;\;\;\;\; \beta_{n,M+1}=+\beta_{n,1}, 
\;\;\; (n=1,2,...,N);
\end{eqnarray*}
\begin{equation}
\lambda_{\hat G}=\prod_{n}\lambda_{{\hat g}_n}=(-1)^N=\prod_{m}\lambda_{{\hat
g}_m}=(-1)^M=-1,
\end{equation}
for  $N$ and $M$ odd. It is obvious that the constraints on parity of $N$ and 
$M$ are not important here, because we can always choose $N$ and $M$ in the 
form $(N=2N'+1, \;\; M=2M'+1)$, and then go to infinity with $N'$ and $M'$ 
independently. 

One can show exactly that the boundary conditions for the $\alpha$- and 
$\beta$- operators chosen this way are not contradictory, if we take into 
account simultaneously the conditions of local equality of the occupation 
numbers for the $\alpha$- and $\beta$- fermions $(5.14)$. As a result we can 
write down the following expressions for the operators $T_{1,2}$ $(5.25-26)$: 
\begin{eqnarray}
T^{\pm}_{1}=\exp\left[K_{1}\sum_{n,m=1}^{N,M}(\beta_{nm}^{\dag}-\beta_{nm})
(\beta^{\dag}_{n+1,m}+\beta_{n+1,m})\right] ,
\end{eqnarray}
\begin{eqnarray}
T^{\pm}_{2}=\exp\left[K_{2}\sum_{n,m=1}^{N,M}(\alpha^{\dag}_{nm}-\alpha_{nm})
(\alpha^{\dag}_{n,m+1}+\alpha_{n,m+1})\right] ,
\end{eqnarray}
where the upper sign $(+)$ corresponds to the states with even numbers of 
fermions $(\lambda_{\hat G}=+1)$, and the lower sign $(-)$ corresponds to the 
states with odd numbers of fermions $(\lambda_{\hat G}=-1)$, with the 
appropriate boundary conditions $(5.41-43)$. This way, the form of the operators 
$T_{1,2}$ for the even and odd states is preserved. What is changing is only 
the boundary conditions. 

\widetext
\section{THE PARTITION FUNCTION}

In this section we perform all the calculations for the statistical sum 
written in the form $(5.36)$, and in the end of the section we only give the 
results for $Z_2(h)$ written in the form $(5.37)$ symmetric in the magnetic 
operator. This can be done almost automatically, since all calculations in 
the latter case are analogous to the ones given below. 

Collecting all the results derived above, we can write the following expression 
for the statistical sum $(5.36)$:
\begin{equation}
Z_2(h)=(2\cosh h)^{NM}<0|T|0>, \;\;\;\; T\equiv T^{\pm}_2T^{\pm}_1T^{*}_h\;,
\end{equation}
where the operators $T^{*}_h$ and $T^{\pm}_{1,2}$ are defined by the formulas 
$(5.33)$ and $(5.44-45)$. In the formula $(5.33)$ one should only omit the 
factor $\exp(\alpha\sum_{nm}\beta^{\dag}_{nm})$, and move the constant factor
$(\cosh h)^{NM}$ out of the expression $<0|(...)|0>$. Let us remind before 
the diagonalization of the $T$- operator in $(6.1)$, in which the multiplicative 
components $T^{\pm}_{1,2}$ and $T^*_h$ are expressed by the Fermi operators of 
the $\alpha$- and $\beta$- types, that these operators satisfy the mixed 
commutative relations $(5.17-19)$. As a result also their Fourier transforms 
will satisfy in general rather complex commutative relations. 

\subsection{Momentum Representation}

Let us pass now to the momentum representation:
\begin{equation}
\alpha^{\dag}_{nm}=\frac{\exp(i\pi/4)}{(NM)^{1/2}}\sum_{q,p}e^{-i(nq+mp)}
\xi^{\dag}_{qp},\;\;\;\;\; \beta^{\dag}_{nm}=\frac{\exp(i\pi/4)}{(NM)^{1/2}}
\sum_{q,p}e^{-i(nq+mp)}\eta^{\dag}_{qp},
\end{equation}
where the factor $\exp(i\pi/4)$ was introduced for convenience. Antiperiodic  
boundary conditions give: $e^{iqN}=-1, \;\; e^{ipM}=-1$, where
\begin{equation}
q(p)=\;\;{\pm}\frac{\pi}{N(M)}, \;\;\;\;\pm\frac{3\pi}{N(M)}, \;\;\;\; 
\pm\cdots,
\end{equation}
and the periodic boundary conditions give: $e^{iqN}=1, \;\; e^{ipM}=1$, where
\begin{equation}
q(p)=0,\;\;\;\;{\pm}\frac{2\pi}{N(M)}, \;\;\;\;\pm\frac{4\pi}{N(M)}, \;\;\;\;
 \pm\cdots.
\end{equation}
Substituting $(6.2)$ into $(5.33)$ and $(5.44-45)$ we get after straight 
forward transformations the following expressions for the operators 
$T^{\pm}_{1,2}$ i $T^{*}_h$:
\begin{eqnarray}
T^{\pm}_1=\exp\left\{2K_1\sum_{0\leq{q,p}\leq\pi}\left[(\eta^{\dag}_{qp}\eta_{qp}+
\eta^{\dag}_{q,-p}\eta_{q,-p}+\eta^{\dag}_{-qp}\eta_{-qp}+\eta^{\dag}_{-q,-p}
\eta_{-q,-p})\cos q+\right.\right.\nonumber\\
\left.\left.(\eta^{\dag}_{-q,-p}\eta^{\dag}_{q,p}+\eta^{\dag}_{-q,p}\eta^{\dag}_{q,-p}+
\eta_{q,p}\eta_{-q,-p}+\eta_{q,-p}\eta_{-q,p})\sin q\right]\right\}=
\prod_{0\leq{q,p}\leq\pi}T^{\pm}_1(q,p),
\end{eqnarray}
\begin{eqnarray}
T^{\pm}_2=\exp\left\{2K_2\sum_{0\leq{q,p}\leq\pi}\left[(\xi^{\dag}_{qp}\xi_{qp}+
\xi^{\dag}_{q,-p}\xi_{q,-p}+\xi^{\dag}_{-qp}\xi_{-qp}+\xi^{\dag}_{-q,-p}
\xi_{-q,-p})\cos p+\right.\right.\nonumber\\
\left.\left.(\xi^{\dag}_{-q,-p}\xi^{\dag}_{q,p}+\xi^{\dag}_{q,-p}\xi^{\dag}_{-q,p}+
\xi_{q,p}\xi_{-q,-p}+\xi_{-q,p}\xi_{q,-p})\sin p\right]\right\}=
\prod_{0\leq{q,p}\leq\pi}T^{\pm}_2(q,p),
\end{eqnarray}
\begin{equation}
T^{*}_h=\exp\left\{\sum_{0\leq{q,p}\leq\pi}\left[\alpha(h,q)(\eta^{\dag}_{-q,-p}
\eta^{\dag}_{q,p}+\eta^{\dag}_{-q,p}\eta^{\dag}_{q,-p})\right]+\Phi(h)\right\}=
\prod_{0\leq{q,p}\leq\pi}T^{\pm}_h(q,p), 
\end{equation}
where $\alpha(h,q)\equiv\tanh^{2}h\frac{1+\cos q}{\sin q}$. 

In the formulas $(6.5-7)$ the upper sign $(+)$ corresponds to the case of even 
states, for which one should omit the term $\Phi(h)$ in the formula $(6.7)$, 
and the lower sign $(-)$ corresponds to the case of odd states with respect 
to the operator of the total number of particles $(\hat S)$. The function 
$\Phi(h)$ is of arbitrary complicated form and we will not write it down here. 
We only mention that it plays an analogous role to the role played by the 
function $f(q)$ in the one-dimensional case $(3.11)$. We used commutativity 
of the operators $T^{\pm}_{1,2}(q,p)$ and $T^{\pm}_h(q,p)$ for different 
$(q,p)$ to write the expressions $(6.5-7)$. Namely, 
\begin{eqnarray*}
[T^{\pm}_1(q,p),T^{\pm}_1(q',p')]_{-}=[T^{\pm}_2(q,p),T^{\pm}_2(q',p')]_{-}=
[T^{\pm}_h(q,p),T^{\pm}_h(q',p')]_{-}=0,
\end{eqnarray*}
as can be easily verified. The statistical sum $(6.1)$ we rewrite now in the 
form:
\begin{equation}
Z^{\pm}_2(h)=(2\cosh h)^{NM}<0|\left[\prod_{0\leq{q,p}\leq\pi}T^{\pm}_2(q,p)
\right]\left[\prod_{0\leq{q,p}\leq\pi}T^{\pm}_1(q,p)T^{\pm}_h(,p)\right]|0>,
\end{equation}
where $|0>$ - the function of the fermionic vacuum in the space of occupation 
numbers in the momentum representation. This function was denoted in the same 
way as in the "coordinate" representation but this should not lead to any 
misunderstandings. We used also commutativity of the operators
$T^{\pm}_1(q,p)$ and $T^{\pm}_h(q,p)$ for $(q,p)\neq(q',p')$  to write the 
formula $(6.8)$. The operators $(\xi_{qp},\xi^{\dag}_{qp})$ and $(\eta_{qp},\eta^{dag}_{qp})$ 
satisfy the standard commutation relations. On the other hand the commutation 
relations mixing them are of rather complex form, in contrast to the relations 
in the "coordinate" representation $(5.17-19)$. This, by the way, is the cause 
of the lack of commutativity of the operators $T^{\pm}_2(q,p)$ and
$T^{\pm}_1(q,p)T^{\pm}_h(q,p)$ for $(q,p)\neq(q',p')$. Now we will maximally 
simplify the bra-vector $<0|(...)$ and the ket-vector $(...)|0>$, which are 
present in the expression $(6.8)$ for $T^{\pm}_2(h)$, "transferring" the 
corresponding operators through the vacuum state.

Now, we will consider in some details the case corresponding to the even 
number of fermions $(\lambda_{\hat G}=+1)$ which means the choice of the 
antiperiodic boundary conditions $(6.3)$. In the end of the paper we will 
shortly consider the case of the odd states $(\lambda_{\hat G}=-1)$ to which 
correspond the periodic boundary conditions $(6.4)$. First, let us note that 
the following equality holds
\begin{equation}
\sum_{q,p}\xi^{\dag}_{qp}\xi_{qp}\;\;=\;\;\sum_{q,p}\eta^{\dag}_{qp}\eta_{qp}\;.
\end{equation}
Further, it is obvious that for fixed $(q,p)$ the quantities $T^{\pm}_2(q,p)$ 
and $T^{\pm}_1(q,p)T^{\pm}_h(q,p)$ are represented by the matrices of the 
size $(16\times 16)$, each of which is considered in its own space of states 
${\cal P}_{\xi}$ and ${\cal P}_{\eta}$ , respectively. After introduction of 
the bases, each of which is built of $16$ functions:
\begin{eqnarray}
\Phi_{0}\equiv |0>_{\xi},\;\;\;\;\;\Phi_{q,p}=\xi^{\dag}_{qp}\Phi_{0},\;\;\;\;\;
\Phi_{-q,-p;\;q,p}=\xi^{\dag}_{-q,-p}\xi^{\dag}_{qp}\Phi_{0},\;\; \cdots \\
\Psi_{0}\equiv |0>_{\eta},\;\;\;\;\;\Psi_{q,p}=\eta^{\dag}_{qp}\Psi_{0},\;\;\;\;\;
\Psi_{-q,-p;\;q,p}=\eta^{\dag}_{-q,-p}\eta^{\dag}_{qp}\Psi_{0},\;\; \cdots ,
\end{eqnarray}
where $\Phi_{0}$ and $\Psi_{0}$ are the functions of Fermi vacuum (which, as 
was mentioned above, we denoted by $\Phi_{0}=\Psi_{0}=|0>$), we obtain after 
a sequence of transformations the expression for the statistical sum $Z^+_2(h)$ 
in the case of the even states:
\begin{equation}
Z^{+}_2(h)=(2\cosh h)^{NM}\left(\prod_{0<{q,p}<\pi}A^2_1(q,h)\right)
\left(\prod_{0<{q,p}<\pi}A^2_2(p)\right)<0|{\tilde{T}}^{+}_{2}
{\tilde{T}}^{+}_1(h)|0>,
\end{equation} 
where
\begin{eqnarray}
{\tilde{T}}^{+}_1(h)=\exp\left[\sum_{0<{q,p}<\pi}B_1(q)(\eta^{\dag}_{-q,-p}
\eta^{\dag}_{q,p}+\eta^{\dag}_{-q,p}\eta^{\dag}_{q,-p})\right], \\
{\tilde{T}}^{+}_2=\exp\left[\sum_{0<{q,p}<\pi}B_2(p)(\xi_{q,p}
\xi_{-q,-p}+\xi_{-q,p}\xi_{q,-p})\right],
\end{eqnarray}
and the quantities $A_{1}(q,h)$, ... , $B_2(p)$ are defined by the formulas:
\begin{eqnarray}
A_1(q,h)=\cosh 2K_1-\sinh 2K_1\cos q+\alpha(h)\sinh 2K_1\sin q,\nonumber\\
A_2(p)=\cosh 2K_2-\sinh 2K_2\cos p,\;\;\;\;\alpha(h)=\tanh^2h\frac{1+\cos q}
{\sin q}, \nonumber\\
B_1(q)=\frac{\alpha(h)[\cosh 2K_1+\sinh 2K_1\cos q]+\sinh 2K_1\sin q}
{A_1(q,h)},\;\;\;B_2(p)=\frac{\sinh 2K_2\sin p}{A_2(p)}.
\end{eqnarray}

Analogously, we can after a sequence of transformations, similar to given 
above, get the following expression of the statistical sum understood in the 
form $(5.37)$ symmetrical with respect to the parameter of the external 
magnetic field $h$:
\begin{equation}
Z^{+}_2(h)=(2\cosh^2\frac{h}{2})^{NM}\left(\prod_{0<{q,p}<\pi}C^2_1(q)
\right)\left(\prod_{0<{q,p}<\pi}C^2_2(p)\right)<0|V^{+}_{2}(h/2)
V^{+}_1(h/2)|0>,
\end{equation}
where
\begin{eqnarray}
V^{+}_1(h/2)=\exp\left[\sum_{0<{q,p}<\pi}D_1(q,h/2)(\eta^{\dag}_{-q,-p}
\eta^{\dag}_{q,p}+\eta^{\dag}_{-q,p}\eta^{\dag}_{q,-p})\right], \\
V^{+}_2(h/2)=\exp\left[\sum_{0<{q,p}<\pi}D_2(p,h/2)(\xi_{q,p}
\xi_{-q,-p}+\xi_{-q,p}\xi_{q,-p})\right],
\end{eqnarray}
and the quantities $C_1(q,h/2)$, ... , $D_2(p,h/2)$ are defined by the formulas: 
\begin{eqnarray}
C_1(q,h/2)=\cosh 2K_1-\sinh 2K_1\cos q+\alpha(h,q)\sinh 2K_1\sin q,\nonumber\\
C_2(p,h/2)=\cosh 2K_2-\sinh 2K_2\cos p+\alpha(h,p)\sinh 2K_2\sin p,\nonumber\\
D_1(q,h/2)=\frac{\alpha(h,q)[\cosh 2K_1+\sinh 2K_1\cos q]+\sinh 2K_1\sin q}
{C_1(q,h/2)},\nonumber\\
D_2(p,h/2)=\frac{\alpha(h,p)[\cosh 2K_2+\sinh 2K_2\cos p]+\sinh 2K_2\sin p}
{C_2(p,h/2)},
\end{eqnarray}
where $\alpha(h,x)=\tanh^2(h/2)(1+\cos x)/(\sin x)$. For the purpose of 
derivation of the expressions $(6.16-19)$ we applied the fact for the even 
states the operator $U_2$ in $(5.37)$ gives vanishing contribution to the 
statistical sum $Z^+_2(h)$. We gave here two representations $(6.12)$ and 
$(6.16)$ for $Z^+_2(h)$, because as can be shown they both can be applied in 
the graph theory $\cite{koch95}$ as we mentioned above. 

In principle, one could now expand the vacuum matrix element in $(6.12)$ or 
in $(6.16)$ into a sum of vacuum matrix elements of the type
\begin{eqnarray*}
<0|\xi_{q,p}\xi_{-q,-p}\cdots\eta^{\dag}_{-q',-p'}\eta^{\dag}_{q',p'}\cdots|0>,
\end{eqnarray*}
derive appropriate commutation relations for the $\xi$- and $\eta^{\dag}$-operators, 
and, finally, sum up the series. Nevertheless in practise this task seems to 
be extremely difficult, as believes the author. Therefore we will proceed the 
other way. Namely, we will come back to the "coordinate" representation, i.e. 
to the $\alpha$- and $\beta$- operators. Then the operators ${\tilde{T}}^{+}_{1,2}$ $(6.13-14)$ 
or $V^{+}_{1,2}$ $(6.17-18)$ are expressed as follows:
\begin{eqnarray}
V^{+}_1=\exp\left[\sum^{N}_{n=1}\sum^{N-n}_{l=1}\sum^{M}_{m=1}a(l)
\beta^{\dag}_{nm}\beta^{\dag}_{n+l,m}\right],\nonumber\\
V^{+}_2=\exp\left[\sum^{N}_{n=1}\sum^{M}_{m=1}\sum^{M-m}_{k=1}b(k)
\alpha_{n,m+k}\alpha_{nm}\right],
\end{eqnarray}
where $a(l)$ and $b(k)$ are given by:
\begin{equation}
a(l)=\frac{1}{N}\sum_{0<{q}<\pi}2D_1(q)\sin(lq), \;\;
b(k)=\frac{1}{M}\sum_{0<{p}<\pi}2D_2(p)\sin(kp),
\end{equation}
for the "symmetric" case, and by:
\begin{equation}
c(l)=\frac{1}{N}\sum_{0<{q}<\pi}2B_1(q)\sin(lq), \;\;
d(k)=\frac{1}{M}\sum_{0<{p}<\pi}2B_2(p)\sin(kp),
\end{equation}
for the "nonsymmetric" case, where the quantities $B_1(q)$, ... , $D_2(p)$ 
are defined above by $(6.15)$ and $(6.19)$. Here we used in both cases the 
same notation $V^{+}_{1,2}$, and further we continue with this convention. 
As can be seen from $(6.13-14)$ and $(6.17-18)$, the structure of the 
operators ${\tilde{T}}^{+}_{1,2}$ in the "coordinate" representation is the 
same as in the case $(6.20)$. The only change concerns the weight factors:
$a(l) \rightarrow c(l)$ i $b(k) \rightarrow d(k)$. The whole procedure used 
above corresponds to the renormalization of the interaction constants in the 
former expression $(5.31)$ for the statistical sum. We will be exploring this 
topic more thoroughly in the following papers of this series. Moreover, here 
appears also a delicate problem of the boundary conditions, connected with 
the expressions $(6.20)$. The discussion of this problem we also postpone to 
a future publication. Here we mention only that in the thermodynamic limit we 
can neglect the boundary effects. On the other hand, in the situation at 
hands it is much easier and more convenient to consider the diagram 
representation for the vacuum matrix element $<0|V^{+}_2V^{+}_1|0>$ in the 
"coordinate" representation than in the "momentum" one, which we denote here 
by $S$, i.e. 
\begin{equation}
S=<0|V^{+}_2V^{+}_1|0>\equiv<0|G|0> .
\end{equation}

\subsection{The Diagram Representation for $S$}

Our aim now is to calculate the vacuum matrix element $S$ $(6.23)$ for the 
sum of products of Fermi creation and annihilation operators. The operator 
$G$ entering $(6.23)$ is a polynomial in the variables $a(l)$, $b(k)$, 
$\alpha_{nm}$ and $\beta^{\dag}_{nm}$. Since $G$ enters in the $(6.23)$  
expectation value form $<0|G|0>$, not all terms in the polynomial give a 
different from zero contribution to the matrix element $S$. Expanding $G$ 
and substituting the expansion into $(6.23)$, the quantity $S$ can be 
represented in the form of the sum of the vacuum matrix elements $\sum_{\nu}S_{\nu}$, 
where $S_{\nu}$ is the vacuum matrix element for the $\nu$-th term of the 
polynomial $G$. As it follows from $(6.20)$, all terms of the polynomial $G$ 
are products of various pairs $b(k)\alpha_{n,m+k}\alpha_{nm}$ and 
$a(l)\beta^{\dag}_{nm}\beta^{\dag}_{n+l,m}$, which we will call below $\alpha$- 
pairs and $\beta$- pairs. Obviously, all the terms in the polynomial $G$ with 
non-equal numbers of the $\alpha$- and $\beta$- pairs give vanishing 
contribution. Moreover, not all terms in the polynomial $G$ with equal numbers 
of the $\alpha$- and $\beta$- pairs will give nonvanishing contribution to 
$S$. Namely, the non-zero contribution to $S$ will give only these terms with 
equal numbers of the $\alpha$- and $\beta$- pairs, in which each annihilation 
operator ${\alpha}_{nm}$ is paired with the corresponding creation operator 
$\beta^{\dag}_{n'm'}$ with identical indices $(n=n',m=m')$. In the opposite case 
this term gives obviously no contribution to $S$. 

This way we arrive at a diagrammatic representation by noticing that to each 
vacuum matrix element $S_{\nu}$ we can uniquely assign a set of lines (links), 
connecting some of the sites of the lattice. For example, to the graphs at 
$Fig.1,a) - d)$ correspond the following matrix elements:
\begin{eqnarray}
a)\;\;\;a(2)b(3)<0|\alpha_{n,m+3}\alpha_{nm}\beta^{\dag}_{n,m+3}\beta^{\dag}_{n+2,
m+3}|0>\;;\nonumber\\
b)\;\;\;a^{2}(1)a(2)b^{2}(1)b(2)<0|\alpha_{n,m+1}\alpha_{nm}\alpha_{n+1,m+1}
\alpha_{n+1,m-1}\alpha_{n+2,m}\alpha_{n+2,m-1}\nonumber\\
\times\beta^{\dag}_{n+1,m-1}\beta^{\dag}_{n+2,m-1}\beta^{\dag}_{nm}\beta^{\dag}_{n+2,m}
\beta^{\dag}_{n,m+1}\beta^{\dag}_{n+1,m+1}|0>\;;\nonumber\\
c)\;\;\;a^{2}(1)a^{2}(4)b^{2}(1)b^{2}(2)<0|\alpha_{n,m+1}\alpha_{nm}\alpha_
{n+1,m+1}\alpha_{n+1,m}\alpha_{n+1,m-2}\alpha_{n+1,m-4}\alpha_{n+5,m-2}
\alpha_{n+5,m-4}\nonumber\\
\times\beta^{\dag}_{n+1,m-4}\beta^{\dag}_{n+5,m-4}\beta^{\dag}_{n+1,m-2}\beta^{\dag}_{n+5,
m-2}\beta^{\dag}_{nm}\beta^{\dag}_{n+1,m}\beta^{\dag}_{n,m+1}\beta^{\dag}_{n+1,m+1}|0>\;;
\nonumber\\
d)\;\;\;a^{2}(2)a(4)b(2)b(3)b(5)<0|\alpha_{n,m+2}\alpha_{nm}\alpha_{n+2,m}
\alpha_{n+2,m-3}\alpha_{n+4,m+2}\alpha_{n+4,m-3}\nonumber\\
\times\beta^{\dag}_{n+2,m-3}\beta^{\dag}_{n+4,m-3}\beta^{\dag}_{nm}\beta^{\dag}_{n+2,m}
\beta^{\dag}_{n,m+2}\beta^{\dag}_{n+4,m+2}|0>\;.
\end{eqnarray}
As one can see from the formulas $(6.20)$ and $(6.24)$, to each horizontal 
line of the "length" $k$ corresponds the factor $b(k)$. Also, to each vertical 
line of the "length" $l$ corresponds the factor $a(l)$. The $a(l)$ and $b(k)$ 
are defined by the expressions $(6.22)$ for the nonsymmetric case. As was 
shown above a nonzero contribution to $S$ give only these matrix elements 
$S_{\nu}$, which do not contain equal numbers of the $\alpha$- and $\beta$- 
pairs. Moreover, the necessary condition for a non-zero contribution is the 
annihilation operators $\alpha_{nm}$ pair with the corresponding creation 
operators $\beta^{\dag}_{nm}$. Geometrically this condition means that from the 
whole family of possible graphs only those for which in each site meet under 
"right angle" only zero or $2$ lines (links) give a non-zero contribution to 
$S$. In other words, the graphs in any site of which meet two horizontal or 
two vertical lines are forbidden. The simplest examples of such graphs are 
shown in $Fig.1, b)-e)$. As a result all graphs giving non-vanishing contribution 
to$S$ should be closed. Moreover, in each site of the graphs selfintersections 
of lines (links) are forbidden, since $(\alpha_{nm})^2=(\beta^{\dag}_{nm})^2=0$. 
>From the point of view of the graph theory to the closed graphs described 
above correspond nonoriented Hamilton cycles (with valency of sites $\delta=0,2$) 
on the simple rectangular lattice $\cite{graph67,har73,tutt84}$. 

This way the vacuum matrix element $S$ $(6.23)$ can be represented in the form 
\begin{equation}
S=\sum_{\nu}S_{\nu}=\sum[sum\;\;on\;\;all\;\;closed\;\;graphs],
\end{equation}
where in the calculations every multiple-connected graph is counted as one 
(for example, the graph in the $Fig.1,c)$). Every closed graph gives the 
contribution equal to 
\begin{equation}
(\pm 1)\prod^{s}_{j=1}a(l_j)b(k_j),
\end{equation}
where $s$ is the number of the horizontal links, which is equal to the vertical 
links. Further, applying the connection between the $\alpha$- and $\beta$- 
operators $(5.15-19)$, and the Wick theorem $\cite{wick50,glimm81}$, one can 
show that any vacuum matrix element giving non-zero contribution into the sum 
$S$ $(6.25)$, can be split into a product of the matrix elements, corresponding 
to the connected parts of the graph (which we will call below for shortness 
the simple loops without selfintersections in the sites of the lattice). We 
can check by direct computation, using the commutation relations $(5.17-19)$ 
for the $\alpha$- and $\beta$- operators, that for example the graphs from the 
$Fig.1,(b-d)$ contribute with the sign $(+)$. Other graphs can contribute 
with the sign $(-)$ as well as, for example, the graph in $Fig.1,e$. 
Commutation relations for the $\alpha$- and $\beta$- operators $(5.17-19)$ are 
illustrated in an appealing way in the $Fig.2$, where the distinguished 
operator $\alpha_{nm}$ $(*)$ for the fixed site $(nm)$ commutes with the 
$\beta$- operators in the sites $(n'm')$, signed with the cross $(\times)$. 
For all others sites the $\alpha$- and $\beta$- operators anticommute. As a 
result the contribution from each particular graph splits into a product of 
contributions from the simple loops. The contribution from a simple loop with 
$s$ horizontal and $s$ vertical links is equal to:
\begin{displaymath}
{\cal L}_{s}=(\pm 1)\prod^{s}_{j=1}a(l_j)b(k_j)\;.
\end{displaymath}
The expression for $S$ $(6.25)$ is now of the form:
\begin{displaymath}
S=1+\sum_{\{s\}}{\cal L}_s+\sum_{\{s\},\{q\}}{\cal L}_s{\cal L}_q+\ldots\equiv
{\Gamma}^{(h)}(z_1,z_2,y)\;,
\end{displaymath}
where $a(l_j)$ and $b(k_j)$ are functions of $z_1\equiv\tanh K_1$, 
$z_2\equiv\tanh K_2$ and $y\equiv\tanh^{2}(h/2)$  for the symmetric case, and
$y\equiv\tanh^{2}h$ in the case asymmetric with respect to the parameter of 
the external magnetic field $(h)$. A contribution to $(6.28)$ gives besides 
summation over the number of links $s$, also the summation over all lengths 
of these links $\{k\}$ and $\{l\}$, for fixed $s$. As can be easily seen, the 
summation in $(6.28)$ over the lengths of the horizontal $\{k\}$ and vertical 
$\{l\}$ links is performed independently. In the graph theory $\cite{graph67,har73}$ 
the function $(6.28)$ is called the generating function, as we mentioned above, 
introducing for it the notation ${\Gamma}^{(h)}(z_1,z_2,y)$, where the upper 
index $(h)$ means being a member of the set of Hamilton cycles. The problem 
was reduced this way to the summation over all Hamilton cycles with the 
varying length of the step (edge) on the rectangular lattice of the type 
described above. 

Now, let us note that the graph representation of $Z_2(h)$, described above, 
looks similar to the diagrammatic representation for the statistical sum of 
the $2D$ Ising model in the vanishing magnetic field $(h=0)$, (see, e.g., the 
papers $\cite{hurst61,rumer61,vdov64}$). In this case, as is known $\cite{hurst61}$, 
the statistical sum can be represented in the form:
\begin{equation}
Z(K_1,K_2)=(2\cosh K_1\cosh K_2)^{NM}\sum_{\alpha,\beta}g_{\alpha,\beta}
\tanh^{\alpha}K_1\tanh^{\beta}K_2\;,
\end{equation}
where $g_{\alpha,\beta}$ denotes the number of the closed graphs consisting  
of $\beta$ horizontal and $\alpha$ vertical links. Since these links connect 
the closest sites of the square lattice, to each link $\alpha$ is assigned the 
factor (weight) $\tanh K_1$, and to each link $\beta$ is assigned the factor 
$\tanh K_2$. In some sites of the graph a simple selfintersection is possible, 
i.e. in one site of the graph meet zero, two, or four lines. This corresponds 
to the nonoriented Euler cycles of the degree $\delta\leq 4$, $\cite{graph67,tutt84}$. 
In the $Fig.3$ is shown one of the simplest graphs contributing to the sum 
$(6.29)$ for $Z_2(K_1,K_2)$. The essential difference of this case in 
comparison with the case with the field $(h)$, described by us above, lies in 
the latter property, because in our case in one site of the lattice can meet 
only zero or two lines (horizontal and vertical). This corresponds, as was 
discussed above, to the nonoriented Hamilton cycles on the square lattice 
$\cite{graph67,tutt84}$. The second difference is that the $\alpha$- and 
$\beta$- links in $(6.28)$ can connect not only the nearest sites of the 
lattice. This result in the appearance of dependence of the weight factors 
$a(l_j)$ and $b(k_j)$ on the distances $l$ and $k$ between the sites of the 
lattice in the vertical and the horizontal direction, respectively. As we 
mentioned above, the problem of calculation of the sum $(6.28)$ can be called 
in the language of the graph theory $\cite{graph67}$ the problem of summation 
over the Hamilton cycles (simple cycles) on the rectangular lattice with 
$(N\times M)$ sites with varying "length" of the edges in the horizontal and 
in the vertical directions, respectively. Simultaneously, the problem of the 
sum $(6.29)$ is equivalent to the problem of summation over all possible 
Euler cycles, described above of the type $(\delta\leq 4)$ on the same lattice. 
As is known $\cite{graph67}$, there is a close correspondence between the 
Euler and the Hamilton graphs. For some types of the Euler graphs one can 
consider instead the corresponding Hamilton graphs. The reversed statement is 
not true. In the papers $\cite{koch95}$ is shown one more example of 
the nontrivial connection between the generating functions for the Euler 
cycles and the Hamilton cycles on the simple rectangular lattice. Namely, in 
the papers $\cite{koch95}$ was shown that the generating function 
$\Gamma^{(h)}(z_1,z_2,y=0)$ for the Hamilton cycles described above is exactly 
equal to the generating function $\Gamma^{(e)}(z_1,z_2)$ for the Euler cycles 
$(\delta\leq 4)$ for the $2D$ Ising model $\cite{graph67}$. Therefore, the 
following equality is true:
\begin{equation}
\Gamma^{(h)}(z_1,z_2)=\prod^N_{n=1}\prod^M_{m=1}\left[(1\!\!+\!\!z_1^2)
(1\!\!+\!\!z_2^2)\!\!-\!\!
2z_1(1\!\!-\!\!z_2^2)\cos{\frac{2\pi n}{N}}\!\!-\!\!2z_2(1\!\!-\!\!z_1^2)
\cos{\frac{2\pi m}{M}}\right]^{\frac{1}{2}},
\end{equation}
where $z_{1,2}\equiv\tanh K_{1,2}$. Taking in $(6.16)$ the external magnetic 
field to be equal to zero $(h=0)$, and using the equality $(6.30)$ we arrive 
at the classical expression $\cite{onsager44}$ for the free energy on one 
Ising spin in the $2D$ Ising model. Let us note that the contribution of each 
graph (connected or disconnected), which consists of a set of the Hamilton 
cycles, can be represented in the form of a product of the determinants of the 
incidence matrices $B_{\nu}$: 
\begin{eqnarray*}
(\pm)\prod^{\omega}_{\nu}\det|B_{\nu}|\;,
\end{eqnarray*}
where $\omega$ denotes the order of connectedness of the graph under 
consideration. It is equal to the number of the simple loops creating the 
graph. This way we conclude that for the computation of the statistical sum 
for the $2D$ Ising model in the external magnetic field it is necessary to 
calculate the generating functions for the Hamilton graphs on the simple 
rectangular lattice of the type described above (see 
$\cite{koch95},\cite{999en.96},\cite{999r.96}$).

\section{LIMITING CASES}

\subsection{The Onsager solution}

Let us shortly discuss one of the method of receiving the Onsager solution 
$\cite{999r.96}$. Putting in Eqs.$(6.16)$ and $(6.19)$  the magnetic field 
equal to zero $(h=0)$, the partition function $Z_2$  $(6.16)$ takes the form:
\begin{equation}
Z_{2}=2^{NM}[(1-z_1^{2})(1-z_2^{2})]^{-\frac{NM}{2}}<0|T^{*}_2T^{*}_1|0>,
\end{equation}
where $z_{1,2}=\tanh {(K_{1,2})}$, while operators $T^{*}_{1,2}$ can be 
written in the "coordinate" representation as:
\begin{equation}
T^{*}_1=\exp\left\{\sum_{n=1}^N\sum_{m=1}^M\sum_{l=1}^{N-n}z_{1}^l
\beta^{\dag}_{nm}\beta^{\dag}_{n+l,m}\right\}, \;\;\;\;\;\;\;
T^{*}_2=\exp\left\{\sum_{n=1}^N\sum_{m=1}^M\sum_{k=1}^{M-m}z_{2}^k
\alpha_{n,m+k}\alpha_{nm}\right\}.
\end{equation}

The Ward--Kac solution $\cite{kw52}$, briefly described in $\cite{feyn72}$, 
contains a topological considerations. Namely, for a given closed graph (we 
consider here Euler graphs on a lattice) a factor $\alpha=\exp(i\pi/4)$ is 
added to a left turn, and a factor $\alpha^{-1}=\exp(-i\pi/4)$ to a right 
turn. Closed graphs (i.e. which we want to include) are  thus taken into account 
and forbidden graphs are compensated if me follow various paths over these 
graphs. Full proof of this theorem was given by Sherman $\cite{sher60}$. 
Similar result holds for hamiltonian graphs on a lattice with variable length 
described above which will be shown in simple cases below. However, we will 
follow the methods of $\cite{vdov64,land76}$ in our consideration. 

First of all let us mention that some of hamiltonian loops (e.g. $Fig.1,e$) 
contribute with minus $(-)$ sign in formula $(6.28)$ for $S$. Namely 
straightforward verification, with the help of commutation relations $(5.17-18)$ 
shows that each doubly intersecting link of the one shown in $Fig.1,e$ 
contributes a minus sign to a overall sign of a simple loop $(6.27)$ for all 
admissible diagrams. At the same time each "simple double link" of the type 
shown in $Fig.1,f$ contributes a $(+)$ sign to the overall sign of a simple 
loop $(6.27)$. All other simple loops without "double links" of the shown in 
$Fig.1,b--d$ come with a plus $(+)$ sign in the sum $(6.28)$. (Let us note 
that there is a one to one correspondence between Euler graphs on a lattice 
and hamiltonian graphs with variable step without "double links", the hamiltonian 
graph may contain one, two or more simple loops. In order to establish this 
correspondence it is necessary to select in the Euler graph all intermediate 
vertices together with intersecting horizontal and vertical links of the Euler 
graph.) 

It is easy to understand now, that if in expression $(6.28)$ for $S$ all simple 
loops are taken with a $(+)$ sign, all left (and right) turns in a simple loop 
give a factor $\alpha=\exp(i\pi/4)$,  ($\alpha^{-1}=\exp(-i\pi/4)$), than the 
problem of calculating the sum for $S$ $(6.28)$ is in fact reduced to a 
"random walk" on a lattice with variable step $\cite{vdov64,feyn72,land76}$. 
In fact, with such a way of following simple loops all loops with "double 
links" cancel (e.g. loops in $Fig.1,e$ and $d$), as it should be. In this way 
one can follow all the loops with "double links" and verify that they cancel 
each other. Moreover, one can check using various examples, following the 
same reasoning as given in $\cite{vdov64,feyn72,land76}$ that if we follow 
various paths over all hamiltonian loops with variable step without "double 
link" (including relevant weights $\alpha$ and $\alpha^{-1}$ at each turn) 
than all the allowed diagrams will cancel. One should stress here that such 
full cancellation of forbidden diagrams in every order takes place only in the 
case of factorizable weights $(z^l_1,z^k_2)$ corresponding to step lengths $l$ 
and $k$, respectively. 

Returning to our problem and using the results of $\cite{vdov64,land76}$, we 
obtain for $S$ $(6.28)$ the following expression: 
\begin{equation}
S=\exp[-\sum_{r=1}^{\infty}f_r],
\end{equation}
where $f_r$ - sum over all single loops with length $(r=2s)$, i.e. consisting 
of $s$ horizontal and $s$ vertical links. Each horizontal line contributes a 
factor $(z^k_2e^{i\varphi/2})$, and each vertical line  - a factor $(z^l_1e^{i\varphi/2})$, 
where angle $(\varphi=\pm\pi/2)$ corresponds to left or right turn. Introducing 
quantity $W_r(n,m,\nu)$ - sum over all possible paths with number of links 
equal to $(r=s_1+s_2)$ from a given initial point $(n_0,m_0,\nu_0)$ to a point 
$(n,m,\nu)$, where $\nu$ - auxiliary index, corresponding to four directions 
$(1,2,3,4)$ on a square lattice, we get for $f_r$:
\begin{equation}
f_r=\frac{1}{2r}\sum_{n_0,m_0,\nu_0}W_r(n_0,m_0,\nu_0) .
\end{equation}
One can easily get the following recursion relations for $W_r(n,m,\nu)$ with 
$(\alpha\equiv\exp(i\pi/4))$:
\begin{eqnarray}
W_{r+1}(n,m,1)=0+\alpha^{-1}\sum_{l=1}^Nz_1^lW_r(n-l,m,2)+0+\alpha\sum_{l=1}^N
z_1^lW_r(n+l,m,4);\nonumber \\
W_{r+1}(n,m,2)=\alpha\sum_{k=1}^Mz_2^kW_r(n,m-k,1)+0+\alpha^{-1}\sum_{k=1}^M
z_2^kW_r(n,m+k,3)+0;\nonumber \\
W_{r+1}(n,m,3)=0+\alpha\sum_{l=1}^Nz_1^lW_r(n-l,m,2)+0+\alpha^{-1}\sum_{l=1}^N
z_1^lW_r(n+l,m,4);\nonumber \\ 
W_{r+1}(n,m,4)=\alpha^{-1}\sum_{k=1}^Mz_2^kW_r(n,m-k,1)+0+\alpha\sum_{k=1}^M
z_2^kW_r(n,m+k,3)+0. 
\end{eqnarray}
The meaning of recursion relations $(7.5)$ is evident. Since the point $(n,m,1)$ 
can be reached from $(n^{\prime},m,2)$ and $(n^{\prime\prime},m,4)$; i.e. from 
above and from below (direction $"1"$ was chosen to be "right"), where
$n'=n-l, n^{\prime\prime}=n+l$, and $l$ ranges, strictly speaking, from $1$ to 
$N-1$. However, for large $N$ the summation over $l$ can be extended to $N$, 
which was done in expression $(7.5)$, because in the thermodynamic limit these 
boundary conditions do not play a role. Hamiltonian structure of simple loops 
is evident in the structure of recursion relations $(7.5)$, which should be 
compared to the case of Euler graphs $\cite{feyn72,land76}$. Writing the 
relations $(7.5)$ in matrix form
\begin{equation}
W_{r+1}(n,m,\nu)=\sum_{n',m',\nu^{\prime}}\Lambda(n,m,\nu|n',m',\nu^{\prime})
W_r(n',m',\nu^{\prime}),
\end{equation}
one can easily see that the following relation holds:
\begin{equation}
Tr\Lambda^r=\sum_{n_0,m_0,\nu_{0}}W_r(n_0,m_0,\nu_{0}),
\end{equation}
and also
\begin{equation}
f_r=\frac{1}{2r}Tr\Lambda^r=\frac{1}{2r}\sum_i\lambda_i^r .
\end{equation}
Taking into account $(7.4)$ and $(3.6)$ we get for $S$, $(7.3)$ the following 
relation:
\begin{equation}
S=\prod_i\sqrt{1-\lambda_i},
\end{equation}
where $\lambda_i$ - eigenvalue of the matrix $\Lambda(n,m,\nu)$, $(i=1,2, 
...,4NM)$. The matrix $\Lambda(n,m,\nu)$ can be easily diagonalized over 
indices $(n,m)$ with the help of Fourier transformation:
\begin{equation}
W_r(n,m,\nu)=\sum_{q,p=0}^{N,M}e^{\frac{2\pi i}{N}nq + \frac{2\pi i}{M}mp}
W_r(q,p,\nu).
\end{equation}
Inserting $(7.10)$ into $(7.5)$, for fixed $(q,p)$ we get:
\begin{equation}
\Lambda(q,p,\nu |q,p,\nu^{\prime})= 
\left[
\begin{array}{cccc}
0 & \alpha^{-1}\sum_{l=1}^Nz_1^l\varepsilon^{-lq} & 0 & \alpha\sum_{l=1}^N
z_1^l\varepsilon^{lq} \\
\alpha\sum_{k=1}^Mz_2^k\omega^{-kp} & 0 & \alpha^{-1}\sum_{k=1}^Mz_2^k
\omega^{kp} & 0 \\
0 & \alpha\sum_{l=1}^Nz_1^l\varepsilon^{-lq} & 0 & \alpha^{-1}\sum_{l=1}^N
z_1^l\varepsilon^{lq} \\
\alpha^{-1}\sum_{k=1}^Mz_2^k\omega^{-kp} & 0 & \alpha\sum_{k=1}^Mz_2^k
\omega^{kp} & 0 
\end{array}
\right],
\end{equation}
where $\alpha\equiv\exp(i\pi/4)$, $\varepsilon\equiv\exp(2\pi\*i/N)$, $\omega
\equiv\exp(2\pi\*i/M)$.

It is evident, that for fixed $(q,p)$ it suffices to calculate the determinant 
of $(4\times 4)$ matrix:
\begin{equation}
\prod_{j=1}^4(1-\lambda_j)=Det(\delta_{\nu\nu'}-\Lambda_{\nu\nu'})\equiv 
A(q,p),
\end{equation}
and after simple calculations for $A(q,p)$, $(7.12)$ we get the following 
formula:
\begin{equation}
A(q,p)=\frac{(1+z_1^2)(1+z_2^2)-2z_1(1-z_2^2)\cos(2\pi q/N)-2z_2(1-z_1^2)
\cos(2\pi p/M)}{(1-2z_1\cos(2\pi q/N)+z_1^2)(1-2z_2\cos(2\pi p/M)+z_2^2)}.
\end{equation}
In $(7.13)$ we have neglected the terms proportional to $z_1^N$ and $z_2^M$, 
since for large $N$ and $M$ , $z_1^N\approx 0$ i $z_2^M\approx 0$, for 
$z_{1,2}<1$. Finally for asymptotically large $(N,M)$ for $S$ $(7.9)$ with the 
help of $(7.13)$ we get:
\begin{eqnarray*}
S=\prod_i\sqrt{1-\lambda_i}=\prod_{q,p=0}^{N,M}A^{1/2}(q,p) 
\end{eqnarray*}
\begin{equation}
=\prod_{q,p=0}^{N,M}\left[\frac{(1+z_1^2)(1+z_2^2)-2z_1(1-z_2^2)
\cos(2\pi q/N)-2z_2(1-z_1^2)\cos(2\pi p/M)}{(1-2z_1\cos(2\pi q/N)+z_1^2)(1-
2z_2\cos(2\pi p/M)+z_2^2)}\right]^{1/2}.
\end{equation}
Of course, for asymptotically large $(N,M)$ the expression $(7.14)$ goes to 
expression $(6.30)$, because of following relations 
\begin{eqnarray*}
\prod_{q=0}^N(1-2z_1\cos\frac{2\pi q}{N}+z_1^2)=1, \;\;\;
\prod_{p=0}^M(1-2z_2\cos\frac{2\pi p}{N}+z_2^2)=1,
\end{eqnarray*}
for $(N,M\rightarrow\infty)$, $z_{1,2}<1$. Finally, using $(6.23)$ and 
inserting $(7.14)$ into formula $(7.1)$, for free energy per Ising spin in 
the thermodynamic limit we get the well known Onsager solution 
$\cite{onsager44}$. 
The method of finding the Onsager solution, given in this paper, 
disregarding its complications, allows for analytical study of the 
Ising-Onsager problem in external magnetic field in several limiting cases 
in two and three dimensions. The proposed method of receiving the Onsager 
solution, as well as previously known graphical methods, work only for case
$(a(l)=z_1^l,\;\;b(k)= z^k_2,\;\; l(k)=1,2, ...)$. Can be shown that all 
these methods are not applicable if faktors $(a(l)$ and $b(k))$ have 
different functional structure. Anyhow, contrary to all previously derived 
methods (graphical et al.) the presented method allows, in such or other 
approximation, for accounting external magnetic field $H$.

\subsection{Low-temperature asymptotic for $F_{2D}(h)$}

The aim of this chapter is to consider the free energy per one Ising spin 
in the external magnetic field for some limit cases. For that reason the parameters 
$(K_{1,2},h)$ are to be renormalised in the following way $(K_{1,2}\geq 0)$:
\begin{eqnarray}
\sinh2K^*_{1,2}=\beta_{1,2}[\sinh2K_{1,2}(1-\tanh^2(h/2)],\nonumber \\
\cosh(2K^*_{1,2})=\beta_{1,2}[\cosh2K_{1,2}+\tanh^2(h/2)\sinh2K_{1,2}],
\nonumber \\
\beta_{1,2}=[1+2\tanh^2(h/2)\sinh2K_{1,2}e^{2K_{1,2}}]^{-1/2}, \;\;\; 
\tanh^2h^*_{1,2}=\tanh^2(h/2)\frac{\beta_{1,2}\exp(2K_{1,2})}{\cosh^2K^*_{1,2}}
\end{eqnarray}
The above presented formulae are adequate for symmetrical case. For asymmetric 
case it is sufficient to substitute, for instance, $K^*_2\rightarrow K_2, h^*_2\rightarrow 0$. 
In short, in this case only the parameter $K_1$ and the field $h$ are subjects 
of renormalisation. Formulae $(6.21)$ and $(6.22)$ included in $(6.20)$, 
take the form:
\begin{equation}
a(l)={z^*_1}^l+\tanh^2h^*_1\frac{1-{z^*_1}^l}{(1-z^*_1)^2}, \;\;\; 
b(k)={z^*_2}^k+\tanh^2h^*_2\frac{1-{z^*_2}^k}{(1-z^*_2)^2},
\end{equation}
for symmetrical case and
\begin{equation}
a(l)={z^*_1}^l+\tanh^2h^*\frac{1-{z^*_1}^l}{(1-z^*_1)^2}, \;\;\;\;\;\;
b(k)=z^k_2,
\end{equation}
for asymmetrical one.

Equations $(7.15-17)$ and the way they were derived point on the possibility 
of obtaining series of asymptotics for free energy per one Ising spin for 
$2D$ Ising model in the external magnetic field $(H)$. In paper $\cite{koch95}$ 
has been shown that vacuum matrix element $S=<0|V_2^+V_1^+|0>$, appearing 
in $(6.16)$ for $Z_2^+(h)$, for case $(a(l)=y, \;\;\; b(k)=z_2^k)$ is equal 
to:
\begin{equation}
S=\Gamma^{(h)}(z_2,y)=\prod_{0<q,p<\pi}\left[1+z^{2}_
{2}+2z_{2}y-2z_{2}(1-y)\cos(p)\right]^2,
\end{equation}
The above formula may be used to obtain low-temperature asymptotic solution 
for the free energy $F_{2D}(h)$ per one Ising spin in the thermodynamic 
limits. Note that the condition $[\tanh^2h^*/(1-z^*_1)^2]\rightarrow 1$,  
together with $(7.15)$ is equivalent to $(\exp(-2K_1)(1-\tanh^2h)\rightarrow 0)$. 
For given $J_1=const, H=const$ the above formulated condition is fulfilled 
for temperature area $T$, when $h\sim{\varepsilon}^{-1}, \;\;\;\varepsilon\ll 1$. 
For that reason, if for instance $([1-\tanh^2h^*/(1-z^*_1)^2]\sim\varepsilon$, 
then $a(l)=\tanh^2h^*/(1-z^*_1)^2+\sim\varepsilon{z^*_1}^l$. Consequently 
in this case the result $(7.18)$ may be applied. To prove it let us consider 
Eq. $(6.15)$ for $B_1(q,h)$, expressed by renormalised parameters 
$(h^*,\; K_1^*)$:
\begin{eqnarray}
B_1(q,h)=\frac{\tanh^2h^*\frac{\sin q}{1-\cos q}+2z_1^*\sin q}
{1-2z_1^*\cos q +{z_1^*}^2},
\end{eqnarray}
where $z^*_1=\tanh K^*_1$, a $h^*$ i $K_1^*$ connected with $h$ i $K_1$ 
as was shown in $(7.15)$. Moreover, due to identity
\begin{eqnarray*}
\frac{z_1^*}{1+{z_1^*}^2}=\frac{z_1(1-\tanh^2h)}{1+2z_1\tanh^2h+z^2_1},
\end{eqnarray*}
introducing a small parameter $(1-\tanh h)\sim\varepsilon, \;\;\; \varepsilon\ll 1$, 
and developing $B_1(q,h)$ into series along $\varepsilon$ $(z^*_1\sim\varepsilon)$, 
we obtain 
\begin{eqnarray*}
B_1(q,h)=\frac{(\tanh^2h^*+2z^*_1)\sin q}{1-\cos q}+\sim{\varepsilon}^2
\end{eqnarray*}
Substituting the last expression to $(6.22)$ we come to the formula
\begin{equation}
a(l)=\tanh^2h^*+2z^*_1,
\end{equation}
describing $a(l)$ with exactness to the second power of $\varepsilon$ $(\sim{\varepsilon}^2)$, 
i.e. in this approximation $a(l)$ does not depend on $l$. Finally, substituting 
in $(7.18)$ $y$ for $a(l)$ expressed by $(7.20)$ we receive in the limiting 
case the following expression for the free energy $F_{2D}(h)$:
\begin{eqnarray}
-\beta F_{2D}(h)\asymp\ln(2\cosh K^*_1\cosh K_2\cosh h)+\frac{1}{2\pi}\int^{\pi}_0
\ln[1+z^{2}_{2}+2z_{2}(\tanh^2h^*+2z^*_1)-2z_{2}(1 - \nonumber \\
\tanh^2h^*-2z^*_1)\cos p]dp,
\end{eqnarray} 
where $h^*$ and $K^*_1$ depend on $h$ and $K_1$ according to $(7.15)$. Note 
that the derived approximation $(7.21)$ may be also applied to the case of 
comparably strong magnetic field $(H)$ for which $(1-\tanh h)\sim\varepsilon, 
\;\;\; \varepsilon\ll 1, \;\; (T=const)$. 

\subsection{High-temperature approximation} 

In the range of high temperature we impose 
$(J_{1,2}/k_BT\sim\varepsilon),\;\;\; \varepsilon\ll 1, \;\;\; (J_{1,2}=const,\;\;H=const)$, i.e. 
$z_{1,2}=\tanh K_{1,2}\sim\varepsilon$. In this approximation the bra-vector 
$<0|T_2$, expressed in terms of $\alpha$-operators by $(5.45)$, can be written 
as:
\begin{eqnarray*}
<0|T_2\simeq <0|\exp(z_2\sum^N_{n=1}\sum^M_{m=1}\beta_{n,m+1}\beta_{nm}), 
\end{eqnarray*}
i.e. expresses in terms of $\beta$-operators, multiplying all phase 
coefficients ${\varphi}_{nm}$  $(5.15)$ by bra-vector $<0|$. It allows for 
diagonalisation of the operator $T=T_2T_1T_h^*$ in $(6.1)$ and calculation of 
the vacuum matrix element $<0|T|0>$. We will not consider the expressions for 
the free energy, as the mentioned above approximation seems to be crude 
approximation and are not of the special interest. 

\section{CONCLUSIONS}

The case of infinitely small external magnetic field is very 
interesting $(h\sim\varepsilon, \;\;\; \varepsilon\ll 1, \;\;\; T=const)$.
Because in Eqs. $(6.15)$ and $(7.15)$ the magnetic field $h$ appears in 
$\tanh^2h$ function, the computations should be carried out up to the second 
term $({\varepsilon}^2)$ inclusive. The presented approach allows for respective 
calculations, nevertheless they are long and complicated enough to present 
them in another paper. We should only like to note here a case connected with 
calculations of the free energy for the external magnetic field $H$ asymptotic 
tending to zero, i.e. fulfilling the condition 
$(h\rightarrow 0, \;\;\; N,M\rightarrow \infty)$. Neglecting in Eq. $(7.16)$ 
for $a(l)$ and $b(k)$ terms proportional to $\tanh^2h^*_{1,2}\sim\tanh^2h/2$ 
for $a(l)$ and $b(k)$ we obtain the following asymptotic expressions:
\begin{eqnarray*}
a(l)\asymp {z_1^*}^l, \;\;\;\;\; b(k)\asymp {z_2^*}^k, \;\;\;\;
(h\rightarrow 0,\;\;\;\; T=const).
\end{eqnarray*}
In this case we can automatically derive the expression for the free energy, 
substituting in $(7.14)$ $z_{1,2}$ for $z^*_{1,2}$:
\begin{eqnarray*}
-\beta F_{2D}(h\rightarrow 0)\asymp\ln 2+2\ln(\cosh h/2)+ \nonumber \\
\frac{1}{2\pi^2}\int^{\pi}_0\int^{\pi}_0\ln[
\cosh{2K_1^*}\cosh{2K_2^*}-\sinh{2K_1^*}\cos q-\sinh{2K_2^*}\cos p]dq dp,
\end{eqnarray*}
where $\cosh{2K_{1,2}^*}$ and $\sinh{2K_{1,2}^*}$ are defined by Eqs. $(7.15)$. 
This is the leading asymptotic term and the latter for $(h=0)$ given Onsager 
solution. The procedure is equivalent to considering the asymptotically 
vanishing magnetic field $h$ in the zero-order approximation, which in the 
author's opinion is worth analyzing. 

The above presented approach to the Lenz-Ising-Onsager problem, on the 
example of $1D$ and $2D$ Ising model in the external magnetic field may be 
extended on the $3D$ Ising model in the external magnetic field for the 
purpose of obtaining the low-temperature approximation. All calculations are 
then, in fact, same as the ones leading to Eq. $(7.21)$, apart from details 
connected with dimension of the considered system. The obtained results will 
be a subject of a future paper. 

\acknowledgements

I am grateful to H. Makaruk, R. Owczarek, and  A. Snakowska for 
their assistance in preparation of the final form of this paper.

\begin{figure}
\caption{Examples of some graphs: (a) -- a self-avoiding walk; (b)-- (f) -- 
Hamilton cycles on a rectangular $(N\times M)$ lattice with equal numbers 
of vertices and edges (with varying length of steps).}
\end{figure}
\begin{figure}
\caption{"Geometry" of transposition relations for $\alpha$- and
$\beta$- operators: * -- $\alpha$-operator; $\times$ -- $\beta$-operator.}
\end{figure}
\begin{figure}
\caption{The simplest example of a graph (Euler cycles) giving a contribution 
to the sum over states $Z(K_1,K_2)$.}
\end{figure}


\begin{references}
\bibitem{lenz20}W. Lenz, Phys. ZS., \ {\bf 21}, 613 (1920).
\bibitem{ising25}E. Ising, Zs. Physik, \ {\bf 31}, 253 (1925).
\bibitem{onsager44}L. Onsager, Phys. Rev., \ {\bf 65}, 117 (1944).
\bibitem{onsager49}L. Onsager, Nuovo Cimento(Suppl.), \ {\bf 6}, 261 (1949).
\bibitem{yang52}C.N. Yang, Phys. Rev., \ {\bf 85}, 809 (1952).
\bibitem{sml64}T.D. Schultz, D.C. Mattis, and E.H. Lieb,\ Rev. Mod. 
Phys., \ {\bf 36}, 856 (1964).
\bibitem{baxter82}R.J. Baxter, {\it Exactly Solved Models in Statistical
Mechanics}, Ac.Press, Inc.,(1982).
\bibitem{izyum87}Yu.A. Izyumov, and Yu.N. Skryabin: {\it Statistical Mechanics
of Magnetically Ordered Systems}, Nauka, Moscow, (in russian), (1987).
\bibitem{mccoy-wu73}B. McCoy, and T.T. Wu, {\it Two Dimensional Ising 
Models}, Harvard U. Press. Cambridge, Mass., (1973).
\bibitem{gaudin83}M. Gaudin, {\it La Fonction D'Onde De Bethe}, MASSON, 
Paris (1983).
\bibitem{liggett85}Th.M. Liggett: {\it Interacting Particle Systems}. 
Springer-Verlag, New York-Tokyo, (1985).
\bibitem{mkoch95}M.S. Kochma\'nski, J.Tech.Phys., {\bf 36}, 485 (1995).
\bibitem{ziman79}J.M. Ziman:{\it Models of Disorder},  Univ. Press, 
Cambridge, (1979).
\bibitem{onkauf49}L. Onsager, B. Kaufman: Phys.Rev, {\bf 76}, 1232 (1949).
\bibitem{koch95}M.S. Kochma\'nski, J.Tech.Phys., {\bf 37}, 67 (1996). 
\bibitem{huang63}K. Huang: {\it Statistical Mechanics}, J.Wiley and Sons, 
Inc., New York -London, (1963).
\bibitem{thompson88}C.J.Thompson,{\it Classical Equilibrium Statistical 
Mechanics}, CLARENDON PRESS-OXFORD, (1988).
\bibitem{graph67} {\it Graph Theory and Theoretical Physics}, ed. F. Harary,
Ac.Press., (New York),(1967).
\bibitem{jordan28}P. Jordan, E. Wigner: Z.Physik, {\bf 47}, 631 (1928).
\bibitem{landau89}L.D. Landau, E.M. Lifschitz: Quantum Mechanics, {\bf III}, 4-th
issue, Nauka, Moscow, (in russian), (1989).
\bibitem{wick50}G.C. Wick,   Phys. Rev., {\bf 80}, 268 (1950).
\bibitem{glimm81}J. Glimm, and A. Jaffe, {\it Quantum Physics}, 
Springer-Verlag, New York (1981).
\bibitem{har73}F.Harary, and E.M.Palmer, {\it Graphical Enumeration},
Ac.Press,(New York),(1973).
\bibitem{tutt84}W.T. Tutte, {\it Graph Theory}, Cambridge University Press,
(Cambridge), (1984).
\bibitem{hurst61}C.A. Hurst, and H.S. Green, J.Chem.Phys.,\ {\bf 33}, 
1059 (1961).
\bibitem{rumer61}A.M. Dyhne, and Yu.B. Rumer, \ Us.Phys.Nauk, {\bf 75}, 
101 (1961), (in russian).
\bibitem{vdov64}N.V. Vdovitchenko, Zh.Eksp.Teor.Fiz. {\bf 47}, 715, (1964)
[Sov.Phys. JETP {\bf 20}, 477 (1965)].
\bibitem{999en.96}M.S. Kochma\'nski, J.Tech.Phys., {\bf 38}, 73 (1997).
\bibitem{999r.96}M.S. Kohma\'nski, Zh.Eksp.Teor.Fiz.{\bf 111}, 1717 (1997)
[JETP {\bf 84}, 940 (1997)].
\bibitem{kw52}M. Kac, and J.C. Ward, Phys. Rev.\ {\bf 88}, 1332 (1952).
\bibitem{feyn72}R.P. Feynman, Statistical Mechanics, W. A. Benjamin, Inc., 
Ad. Book Progr. Reading, Massachusetts (1972).
\bibitem{sher60}S. Sherman, Journ. Math. Phys., \ {\bf 1}, 202 (1960); 
{\bf 4}, 1213 (1963).
\bibitem{land76}L.D. Landau, E.M. Lifschitz: Statistical Physics, {\bf Y}, 
Part 1, 3-th issue, Nauka, Moscow, (in Russian), (1976).

\end{references}
\end{document}